%% file: main.tex
\documentclass[acmtog,nonacm]{acmart}

\usepackage{booktabs} %
\usepackage{dblfloatfix}

\usepackage[ruled]{algorithm2e} %

\SetAlFnt{\small}
\SetAlCapFnt{\small}
\SetAlCapNameFnt{\small}
\SetAlCapHSkip{0pt}

\input{commands}

\begin{document}

\title{GradRig: Differentiable Weights for Skinned Gaussian Splat Deformation}

\author{Nina Vesseron}
\email{nina.vesseron@ensae.fr}
\orcid{0009-0009-4468-0881} 
\affiliation{%
  \institution{ENSAE-CREST}
  \country{France}
}
\affiliation{%
  \institution{Adobe}
  \country{France}
}

\author{Elie Michel}
\email{emichel@adobe.com}
\orcid{0000-0002-2147-3427}
\affiliation{%
  \institution{Adobe}
  \country{France}
}

\renewcommand{\shortauthors}{Vesseron and Michel}

\begin{abstract}
Skinned deformation is a common framework to turn a 3D shape from its rest pose into a dynamic pose through the deformation of a coarser kinematic structure, called rig. When applied to a 3D mesh, this rig only needs to displace vertices to deform the polygons that connect them. However, when deforming 3D Gaussian Splats, which do not provide connectivity information, rigidly transforming points is not enough to prevent the creation of holes when stretching shapes. In this paper, we use the spatial gradient of skinning weights to provide a full mesh-free deformation pipeline for Gaussian Splats, that more accurately stretches splats while remaining fully compatible with real-time rendering capabilities, which we demonstrate in a WebGL viewer. We present how we evaluate these gradients when the user creates the rig structure and propose an optional adaptive resampling scheme to split up splats that still produce artifacts.
\end{abstract}



\begin{teaserfigure}
    \includegraphics[width=\textwidth, trim=0 73cm 0 0, clip]{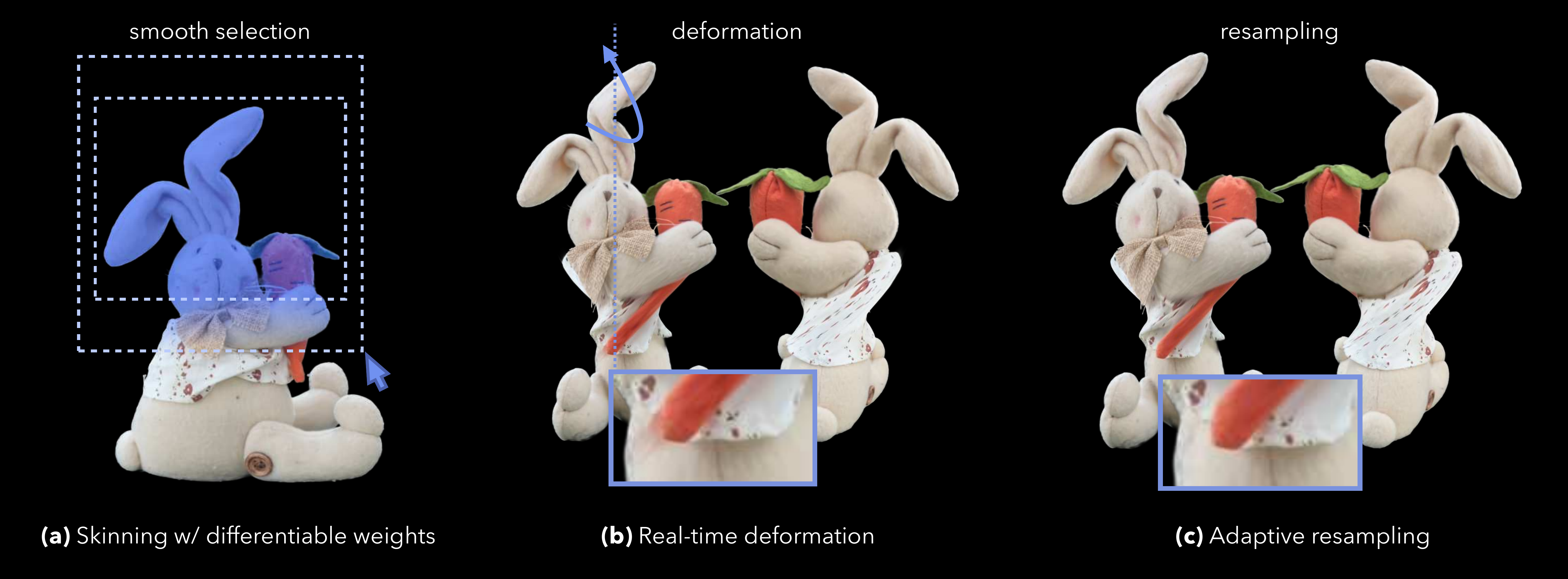}
    \caption{We present a full pipeline to apply skinned deformation to 3D Gaussian Splats (3DGS). When setting up the rig, we evaluate the gradient of skinning weights \textbf{(a)}. These are used during deformation to dynamically stretch splats to prevent holes \textbf{(b)}. To further reduce deformation artifacts, we add an optional adaptive resampling step based on a pretrained resampling dictionary \textbf{(c)}.}
    \Description{TODO}
    \label{fig:teaser}
\end{teaserfigure}

\maketitle

\section{Introduction}

Animating 3D shapes typically consists in starting from a rest shape and deforming it into a specific pose. The pose is slightly different at each frame of the animation, while remaining consistent at the scale of the sequence. Since a shape is usually represented by an aggregation of millions of primitives, it would be both too time consuming and too temporally inconsistent to manually modify every one of them. Instead, computer animation tools provide a means to only deform a coarser structure, and then automatically propagate the deformation to the full shape. This coarse structure can take multiple forms (bounding cage~\cite{Thiery24}, constraint points~\cite{Botsch09}, skeleton~\cite{Weber07}), and we focus here on the most common case, where it represents an internal skeleton of the shape. This is called a rig and often comes with extra constraints about maximum deformation angles, locked transforms, etc. The individual components of a rig are called rig nodes, or sometimes "bones" (for character animation).

The propagation of the deformation from the coarse structure to the full shape relies on an association of the base elements that compose the shape with the rig nodes, which is called "skinning". When the deformed shape is represented as a 3D mesh, each pair of a rig node and a mesh vertex is associated with a skinning weight, which tells how much the vertex is influenced by the pose of the rig node. When the rig pose changes, each vertex is displaced, and since the connectivity information remains unchanged, the polygons smoothly deform over time.

\begin{figure}
    \centering
    \includegraphics[width=\linewidth]{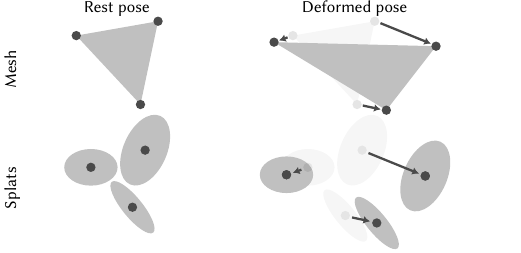}
    \caption{Traditional skinned deformation is applied to each point independently. When deforming a mesh (top row), the surface elements (triangles) are stretched as a consequence of connecting vertices. But when deforming a splat cloud (bottom row), the elements of surface/volumes (splats) are not deformed according to their neighbors because there is no connectivity information. This results in unwanted holes or overlap in the deformed shape. For clarity this example uses rig nodes (not represented in the figure) that are only translated (no scaling/rotation) and each point is influenced by a different mixture of rig nodes.}
    \label{fig:problem-setting}
\end{figure}

In this paper, we apply rigged deformation to 3D shapes represented using 3D Gaussian Splatting (3DGS) rather than a 3D mesh. This representation has shown great flexibility to encode shapes that are not limited to surfaces, yet a key difference is their absence of connectivity information: each point carries its own element of surface/volume without explicit notion of neighboring points. In case of a rigid deformation, where all neighbor points undergo the same deformation, this is not a problem as we can apply the same transform to all splats. However, in regions where neighbor splats follow different deformation -- because they have different skinning weights -- the element of surface/volume carried around each deformed position should account for the stretch or twist introduced by the relative movement of its neighbors. As illustrated in \Fig~\ref{fig:problem-setting}, we address this problem by using a first-order estimation of the neighbors deformation, which we evaluate using the gradient of skinning weights so that it does not depend on the number of neighbors and as such remains easy to parallelize on a GPU.

Our contributions in this paper are:

\begin{itemize}
    \item A lightweight and GPU-friendly skinning technique that deforms Gaussian splat models and automatically stretches Gaussians to prevent the introduction of holes, based on the gradient of skinning weights (\Sec~\ref{sec:grad-rig}).
    \item A framework to define the gradient of skinning weights without explicitly looking at splat's neighbors (\Sec~\ref{sec:diff-weights}).
    \item A refinement post-process that adaptively resamples the deformed Gaussian splats using a scene-agnostic dictionary-based technique. This mitigates the artifacts inherent to parallel per-splat deformation schemes. (\Sec~\ref{sec:resampling})
\end{itemize}

Our method enables real time manipulation of Gaussian splats on consumer-grade device without scene-specific optimization nor mesh extraction. It operates on pre-existing Gaussian splat scenes without access to original images.

\section{Related Work}

\paragraph{Gaussian splatting} While popularized in the context of novel view synthesis, Gaussian splat representations~\cite{kerbl2023gaussians} are increasingly produced by other means such as generative models \cite{ziwen2025longlrm,li2025flashworld,liang2025wonderland}, authoring tools \cite{chen2024GaussianEditor,pandey2025painting}, and simulation pipelines \cite{zhong2024reconstruction}. We thus propose a deformation method that does not assume any particular upstream reconstruction process.

\paragraph{3DGS Animation} Gaussian splats have been used to represent animated scenes, for which temporal variations are often encoded through neural deformation models \cite{wu2024:4dgs,yang2023deformable3dgs,liang2025gaufre,sun2024:3dgstream,luiten2023dynamic} or physically-based constraints \cite{xie2024physgaussian,jiang2024vrgs}. Although compatible with real-time rendering, this implicit or latent representation does not allow user-specified deformations at inference time. Even explicit models~\cite{duan2024:4drotorgs,luiten2023dynamic} are generally tied to the specific motions observed during training. SC-GS~\cite{huang2024scgs} notably enables motion edition thanks to a sparse set of controls, but these require a video input and per-scene training. Our method uses a sparse, explicit and editable representation for Gaussian splat animation that follows the idioms from skeletal animation.

\paragraph{Mesh-based splat deformation} One common solution to enable interactive deformation of Gaussian splats is to rely on a mesh proxy~\cite{huang2024gsdeformer,borycki2025gasp}. For instance, GaussianMesh~\cite{gao2024realtime} connects Gaussians through a mesh that also guides their optimization, although this requires full retraining from images. In these approaches the mesh acts less as the actual rendered shape and more as a deformation structure like a cage or lattice, and these methods are tightly coupled with work that uses splats for the goal of extracting better meshes~\cite{guedon2023sugar,liu2024dgmesh} or to guide optimization~\cite{guedon2025milo}. The main issue however is that meshes cannot adequately handle volumetric or non-surface content such as smoke or foliage. Tetrahedral meshes~\cite{jiang2024vrgs,mai2025radiance} lift this restriction by relying on a solid volume, but at the cost of a substantially heavier preprocessing step.

\paragraph{Deformation refinement} Deforming a Gaussian splat by applying a rigid or affine transform to its mean and covariance is straightforward, but becomes problematic when the deformation field varies significantly across the support of individual splats. One mitigation is to add an anisotropy regularizer during reconstruction that encourages splats to remain near-isotropic~\cite{xie2024physgaussian, feng2025gaussian}. This helps prevent major artifacts under rotations or torsions and corrects the rigid component of the gradient associated with such transformations. However, in the general case, this correction is not sufficient, and it requires a scene-specific retraining pass. We instead introduce a lightweight resampling stage that runs in seconds as a post-process, selectively refining only the splats whose support spans a region of high deformation-field variation. This is complementary to proxy-based deformation pipelines; for instance, our resampling could be plugged into the workflow of~\cite{huang2024gsdeformer} to improve its output quality.

\section{Our method}

Our deformation process is designed around a simple idea. When all splats undergo the same rigid transform, they do not need to be stretched; what drives the deformation of individual splats is indeed the \emph{difference} of transform with respect to their neighbors. In other words, what we need is the \emph{gradient} of the deformation field.

When using skinned deformation, the deformation field is defined by the combination of a few global rigid transforms (the rig nodes) using per-splat skinning weights. In \Sec~\ref{sec:grad-rig} we present a real-time deformation scheme where the overall deformation gradient is obtained from the gradient of individual skinning weights, evaluated at each splat center and stored as extra splat attributes.

We then discuss in \Sec~\ref{sec:diff-weights} how to compute reliable skinning weight gradients. Our key idea is to evaluate these as part of the authoring process that defines skinning weights, so as to avoid discretization artifacts. We do this by instrumenting the authoring tool in a way that does not affect the user workflow.

Our method deforms individual splats in a rigid way, based solely on gradients evaluated at their center. Although this approximation is key to the real-time aspect of our technique, we discuss in \Sec~\ref{sec:resampling} how to overcome the visual artifacts that it may introduce when splats are large with respect to the frequency at which skinning weights vary. We address this using an adaptive resampling scheme that only splits splats that need it, relying on a precomputed dictionary of splat decompositions to speed up the process.

\subsection{Gradient rig for real-time deformation} \label{sec:grad-rig}

In this section, we assume that we know the gradient of skinning weights and describe how they can be used to properly stretch splats during real-time skinned deformation. Gaussian splats are given as a sequence of positions $(p_i)_i$ and a sequence of covariances $(\Sigma_i)_i$, where $i$ is the splat index ($0 \leq i < N$). To enable fast parallel deformation, we consider in this section that the number $N$ of splats does not change. We write respectively $p'_i$ and $\Sigma'_i$ the deformed position and deformed covariance of the splat of index $i$.

The rig consists in $M$ rig nodes and its deformed pose is represented as a sequence $(T'_j)_{0 \leq j < M}$ of rigid transforms. Each transform $T'_j$ is a $4 \times 4$ matrix that encodes a 3D affine map in homogeneous coordinates. To simplify notations we assume here that the rest pose of each node is the identity.
The splats are attached to the rig using skinning weights $w_{i,j} \in \mathbb R$, which give the relative influence that a rig node $j$ has on the splat $i$. We write $W_i = \sum_{j=0}^{M-1} w_{i,j}$ the sum of all weights for a given splat.

\subsubsection{Deformation of $p_i$}

Although more advanced skinning schemes such as dual quaternions~\cite{Kavan07} have been proposed, we demonstrate our method in the case of regular linear blend skinning in order to better highlight the simplicity of our extension.
The center $p'_i$ of the deformed splat is obtained like the deformed vertices in a skinned mesh. We apply the average %
\ninaUpdate{$F_i$} of the rig node transform matrices $T'_j$ weighted by the splat's skinning weights $w_{i,j}$:

\begin{align}
    & p'_i = F_i p_i \label{eq:deform-pos}
    \quad \text{where } \quad  F_i = \frac{1}{W_i} \sum_{j=0}^{M-1} w_{i,j} T'_j
\end{align}

\subsubsection{Deformation of $\Sigma_i$}

To compute the deformed covariance $\Sigma'_i$, we also assume that for each $j$ the skinning weight $w_{i,j}$ is computed, in rest pose, as a field $w_{i,j} = w_j(p_i)$ and that we have access to the spatial gradients $\grad w_j$ at $p_i$. We write $\grad w_{i,j} = \grad w_j(p_i)$ and $F_i = F(p_i)$ becomes a field. We see in \Sec~\ref{sec:diff-weights} how we get these.%

In theory, the element of geometry encoded by the Gaussian splat could be deformed arbitrarily by the rig, requiring to either bend the primitive or resample it. But since we want the deformed shape to remain a cloud of gaussian splats (to be able to chain deformations), and that each splat gets deformed into a single splat (to enable parallelization through a GPU vertex shader), we only apply a first order (i.e., linear) approximation of the deformation field $F$ to the covariance matrix. We can thus encode the deformation of the covariance $\Sigma_i$ as a $3 \! \times \! 3$ matrix $R_i$, namely $\Sigma'_i = R_i \Sigma_i R_i^T$.
If all neighbors of the splat $i$ undergo the very same deformation as $i$ (i.e., if all skinning weights are the same), we rigidly transform the covariance using the upper-left $3 \! \times \! 3$ part of $F_i$:

\begin{equation}
    R_i^{\text{(rigid)}} = \text{mat3x3}(F_i)
\end{equation}

\begin{figure}
    \centering
    \includegraphics[width=\linewidth]{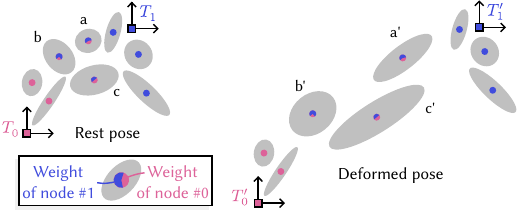}
    \caption{This simplified example highlights why a splat needs to be aware of how its neighbors are deformed in order to properly transform its covariance matrix. Since rig nodes $0$ and $1$ are only translated (no scale nor rotation), the linear blend $F_i$ is also only a translation for any splat $i$. However, splats $a$, $b$ and $c$ need to be elongated in order to properly fill in the gap left by the spatial variation of this translation factor. Instead of costly neighbor lookups, we rely on a precomputed gradient of skinning weights to provide information about neighbor deformation when computing $R_i^{\text(elastic)}$.}
    \label{fig:translation-only}
\end{figure}

However, neighbors may have different skinning weights, resulting in another component in the covariance deformation (see \Fig~\ref{fig:translation-only}).
Formally, this comes from differentiating the field %
\ninaUpdate{$F(p)$} with respect to $p$ (details in supplemental), which leads to a second term where the gradient of skinning weights estimates how the variation of deformation across neighbors results in covariance deformation:

\begin{align}
    & R_i^{\text{(elastic)}} = \sum_{j=0}^{M-1} (T'_j p_i) \,\otimes\, \vec G_{i,j} \\
    & \text{where } \vec G_{i,j} = \frac{\grad w_{i,j}}{W_i} - \frac{w_{i,j}}{W_i^2} \sum_{j'=0}^{M-1} \grad w_{i,j'}
\end{align}

Note that $\vec G_{i,j}$ is the gradient of the normalized weights $w_{i,j} / W_i$, which simplifies into $\grad w_{i,j}$ in cases where all weights sum to 1 (i.e., $W_i = 1$). The outer product $(T'_j p_i) \,\otimes\, \vec G_{i,j}$ is a $3 \! \times \! 3$ matrix.

We combine our rigid and elastic deformation terms with an hyper-parameter $\eta$ that tunes the strength of the elastic term. This $\eta$ would be exactly one if the first order approximation was exact, in practice one can use slightly smaller values to reduce artifacts in previews (our post-processing properly fixes them in \Sec~\ref{sec:resampling}):

\begin{equation} \label{eq:terms-of-R}
    \Sigma'_i = R_i \Sigma_i R_i^T \quad \text{where} \quad R_i = R_i^{\text{(rigid)}} + \eta \, R_i^{\text{(elastic)}}
\end{equation}

\subsubsection{Deformation of view-dependent color}

Many Gaussian splatting flavors allow splats to have a view-dependent color, starting with the original 3DGS paper~\cite{kerbl2023gaussians} which encodes this using Spherical Harmonics (SH). In general, this takes the form of a term $C(\vec d)$, where $\vec d$ is the view direction.
Since we explicitly build a rotation matrix $R_i$ to transform the covariance, we may apply its inverse for the directional color. The color of the splat thus becomes $c_i = C(R_i^T\vec d)$, with $\vec d$ evaluated for the deformed position $p'_i$.

\subsubsection{Implementation}

In practice, we draw splats using GPU accelerated rasterization, where each splat is an instance of a quad. In the vertex shader, the splat's position and covariance are loaded as instance attributes and used to project the 4 corners of the quad.

To implement our gradient-based skinned deformation, we add instance attributes for skinning weights $w_{i,j}$ and their evaluated gradients $\grad w_{i,j} = \grad w_j(p_i)$, and a uniform buffer is used to pass in the $M$ rig node transforms $T'_j$. To limit the memory footprint of the skinning weights, we constrain our scenes to 2 non-zero weights per splat to be able to encode them in a sparse way: we add 2 integer attributes to tell the 2 $j$ indices where $w_{i,j}$ is non zero.

Formulas from equations \ref{eq:deform-pos} to \ref{eq:terms-of-R} are inserted before the computation of corner projection, in which $p'_i$ and $\Sigma'_i$ are then used as drop-in replacements of $p_i$ and $\Sigma_i$ respectively.
When a rig node is transformed, we only need to update the corresponding uniform buffer, and the skinned deformation of the Gaussian splat cloud thus happens in real time.

\subsection{Differentiable skinning weights} \label{sec:diff-weights}

The previous section assumed that we know for each splat $i$ and each rig node $j$ the gradient $\grad w_{i,j} = \grad w_j(p_i)$. When all we know about the field $w_j$ of skinning weight for node $j$ is its value sampled at Gaussian centers, we can only estimate its gradient \textit{a posteriori}, using finite differences with the nearest neighbors of each Gaussian. However, we argue in this paper that we can do better, because in practice we have access to more information: the authoring operations that lead to the definition of skinning weights $w_{i,j}$ can easily be augmented with auto-differentiation. We show that this approach is flexible and captures more accurately the full skinning fields $w_j: \mathbb R^3 \rightarrow \mathbb R$ than looking at nearest neighbors. Nonetheless, our deformation method from \Sec~\ref{sec:grad-rig} can be used with both gradient estimation techniques.

\subsubsection{Selection fields} \label{sec:selection-fields}

We propose to instrument the process that leads to the definition of the weights $w_{i,j}$ so that it automatically provides a gradient as well. When these weights are produced by authoring operations, we consider that users start from $w_{i,j} = 0$ everywhere and then modify the weights using a sequence of smooth screen-space selection followed by a composition operation to \textit{add}, \textit{subtract}, \textit{multiply} or \textit{replace} this selection from the current weights.

\begin{figure}
    \centering
    \includegraphics[width=\linewidth]{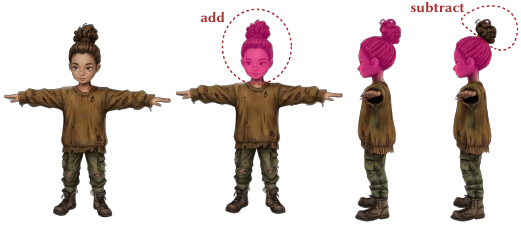}
    \caption{We model the process that defines skinning weights as a sequence of selections that add, subtract or multiply the previous weights. This encompasses the way most authoring tool enable skinning weight definition.}
    \label{fig:skinning-example}
\end{figure}

For instance, \Fig~\ref{fig:skinning-example} shows an example where the user defines the skinning weight for the \textit{head} node of a character rig by drawing a circle around the head, then changes viewpoint, and subtract the hair (that belongs to another node).

Formally, we consider the following steps:

\begin{enumerate}
    \item Pick a viewpoint $V$ (position/orientation/etc.) of the camera.
    \item Draw a selection shape $S$ in screen space.
    \item Set a selection smoothness $\sigma \geq 0$ to define how the weight decays when a splat is close to the edges of the selection.
    \item Set a selection strength $\omega \in (0,1)$.
    \item Set a selection compositing operation $\op$, which is either Replace, Add, Subtract or Multiply (or any other operation that can be auto-differentiated). Return to step (1).
\end{enumerate}

At each iteration of this authoring loop, the skinning weights for the active rig node $j$ are updated as follows:

\begin{align}
    w_{i,j} \leftarrow & \op(w_{i,j}, w'_{i,j}) \\
    & \text{where } w'_{i,j} = \omega \cdot \smoothstep(-\sigma, \sigma, sdf) \\
    & \text{and } sdf = \sdist\!\Big(S, \proj(V, p_i)\Big)
\end{align}

The operator $\sdist(S, x)$ evaluates the signed distance to the selection shape $S$ at a screen space position $x$, and the operator $\proj(V, p)$ computes the screen space position onto which a 3D point $p \in \mathbb R$ gets projected, given the viewpoint information $V$.

This authoring framework enables various means of interaction by playing with the shape $S$ (rectangle, ellipse, free-form contour, brushed shape, etc.) as well as the compositing operation $\op$. A user typically sets the skinning weight field of a rig node in 1 to 5 iterations of the steps described above.

\subsubsection{Auto-differentiation of authoring operations} \label{sec:autodiff-authoring}

In this authoring framework the resulting weights $w_{i,j}$ are entirely defined from expressions that auto-differentiation libraries handle easily, including the signed distance to the screen-space selection shape and the projection (be it perspective or orthographic). As a consequence, implementing this authoring loop with an auto-differentiation library naturally provides the gradients $\grad w_{i,j}$ that are needed in \Sec~\ref{sec:grad-rig}.

\begin{figure}
    \centering
    \includegraphics[width=\linewidth]{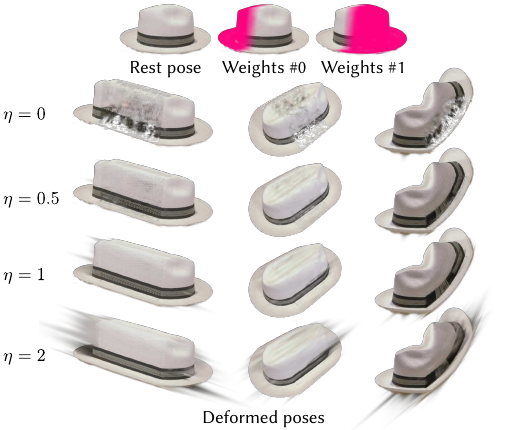}
    \caption{When our gradient rig deformation is not used ($\eta = 0$), holes are introduced because splats are not stretched according to their neighbors deformations. Our gradient rig ($\eta = 1$) addresses this issue, however its linear approximation may overstretch some splats, which shows when artificially doubling the influence of our elastic term ($\eta = 2$). This is solved by our adaptive resampling or by lowering the elastic term (for previsualization).}
    \label{fig:results-hat}
\end{figure}

\subsection{Resampling algorithm} \label{sec:resampling}

As illustrated in \Fig~\ref{fig:results-hat}, the affine approximation used when deforming splats in a parallel pipeline can introduce visual artifacts, particularly under large deformations or when certain splats exhibit large eigenvalues.
We propose to solve this with a post-processing that adaptively splits the splats whose covariance matrix causes artifacts.
We first discuss the root cause of artifacts caused by our deformation model (\Sec~\ref{sec:deformation-model}).
We then introduce the optimization problem used to subdivide a splat (\Sec~\ref{sec:one-splat}), and finally we describe how we construct a scene-agnostic dictionary that speeds-up resampling during (or shortly after) deformation (\Sec~\ref{sec:precomputation}).

\subsubsection{Deformation model} \label{sec:deformation-model}

Our intent to limit the runtime performance overhead introduced by our deformation scheme led to the following assumptions in \Sec~\ref{sec:grad-rig}:
\begin{itemize}
    \item The deformed shape remains within the space of Gaussian splats, i.e., Gaussians that cannot be bent.
    \item Gaussians are small enough to be deformed using an affine approximation of the deformation field $F$: for any point $y$ of the splat, we use $F(y) \approx F(p) + J_F(p) \cdot (y - p)$, where $p$ is the center of the splat and $J_F$ is the Jacobian of $F$.
    \item Gaussians are processed in parallel, and are not aware of their neighbors (except through the gradients $\grad w_j$).
\end{itemize}

The key source of artifact introduced by such a deformation model comes from the affine approximation. When a Gaussian is too large, or when the deformation field $F$ has high frequency variations, the approximation no longer holds. Our solution consists in adaptively splitting Gaussian in the rest shape, given a bound on the error in the deformed shape. \elieUpdate{By operating on the rest shape, this resampling can happen at a slower framerate than deformation, because it remains valid for similar or less intense deformations.}

Formally, in regions of high deformation, we resample splats to enforce that the second-order term of the Taylor expansion remains below an arbitrary threshold~$\varepsilon$, since this term is the leading-order error of the local affine approximation:
\begin{equation} \label{eq:hessian-condition}
    \left\| \sum_{k=1}^3(y - p)^TH_{F_k}(p)(y-p)e_k \right\| < \varepsilon
\end{equation}
where $H_{F_k}$ is the Hessian matrix of the $k$-th dimension of $F$ and $(e_k)_{1\leq k\leq 3}$ is the canonical basis of $\mathbb R^3$. Our approach used in \Sec~\ref{sec:grad-rig} to build the Jacobian matrices can be generalized to get an analytical form for the Hessians as well (see supplemental).

\subsubsection{Subdivision of one splat} \label{sec:one-splat}

Let us consider a single Gaussian, lying at position $p^*$, with covariance $\Sigma^*$ and opacity $o^*$. We intend to replace this single splat by a budget of $L$ new splats, parameterized by $\bar p = (p_l)_l$, $\bar \Sigma = (\Sigma_l)_l$ and $\bar o = (o_l)_l$ ($1 \leq l \leq L$) so that this set of $L$ splats is visually close to the original one while being less prone to artifacts in case of deformation. We obtain these parameters by gradient descent over the following loss:
\begin{equation} \label{eq:loss_resampling}
    \mathcal L = \mathcal L_\text{rendering} + \taureg \, \mathcal L_\text{conditioning}
\end{equation}
where $\taureg$ is a regularization factor.
The first term ensures that at rest pose, our resampling operation does not impact the rendered appearance of the shape. To this extent, we sample the differentiable rendering function $\mathcal R$
from random rays:

\begin{equation} \label{eq:rendering_objective}
    \mathcal L_\text{rendering} = \sum\limits_{\substack{\text{random} \\ \text{ray } r}}\|\mathcal R(r, \bar p, \bar \Sigma, \bar o) - \mathcal R(r, p^*, \Sigma^*, o^*)\|^2
\end{equation}

The second term ensures that the resampling reduces the error of the affine approximation of the deformation field. To do so, we limit $\|y - p\|$ by fostering smaller Gaussian: we penalize the largest eigenvalue across all covariance matrices $\Sigma_l$.
\begin{equation}
    \mathcal L_\text{conditioning} = \softmax\limits_{1 \leq l \leq L}\|\lambda(\Sigma_l)\|_8
\end{equation}
where we approximate the largest eigenvalue of $\Sigma_l$ as the $L_8$ norm of the vector $\lambda(\Sigma_l)$ of its eigenvalues.

\paragraph{Sampling scheme} The evaluation of $\mathcal L_\text{rendering}$ requires to sample random rays, at which we check that drawing only the original splat or only the $L$ new splats yields the same result. Naively drawing rays with respect to a uniform distribution would require a prohibitive amount of samples when the splats are very elongated (e.g. $\lambda_{\max} = 10^6 \lambda_{\min}$). To address this, we pick rays by sampling both its start and end points on the Mahalanobis ellipsis of the Gaussian: we first uniformly sample a position on the unit sphere in a frame aligned with the eigen vectors of the covariance, then scale each dimension by the square root of the corresponding eigenvalue, scaled by a factor 4 to encompass most of the original Gaussian.

\subsubsection{Dictionary-based resampling} \label{sec:precomputation}

The computational cost of generalizing this approach to an arbitrary number of splats would be prohibitive. Instead, we adopt a dictionary-based approach where we first precompute a database of many cases of resampling a single Gaussian splat, then we process the input splats in parallel and look-up this database to quickly produce a global resampling of the whole cloud. This approach is possible because the input space of the database has a low dimension: only 2 eigenvalues of the covariance matrix and the opacity factor matter for the lookup operation. The dictionary is only built once and then used for any scene.

\paragraph{Discretization} We precompute a table for many shapes of Gaussian splats. By symmetry, we ignore the orientation of the splat (we can simply rotate the resampled version) and divide all eigenvalues by the smallest one (we can always scale the resampled version), so that the covariance is $(\lambda_1, \lambda_2, 1)$. We then discretize the scale of possible eigenvalues. The range of eigenvalues is typically between $1$ and $10^4$, but can become very large (see supplemental). We note that the good solution for resampling a splat does not vary much in extreme conditioning, so we can afford less examples in our dictionary for these cases. As a consequence, we use a logarithmic discretization for $\lambda_1$ and $\lambda_2$. More exactly, we consider 40 values between $1$ and $10^4$ on a log scale. We also discretize the scale of opacity $o$, in linear space this time, with 10 increments between 0 and 1. We thus build a dictionary of 16k entries. We use $L = 5$, which yields to a dictionary of about 0.74 MB (using 32 bit float values, no compression). %

\paragraph{Lookup} Once the dictionary is filled with one resampling per entry using the loss described in \Sec~\ref{sec:one-splat}, we can use it to process a whole cloud of Gaussian splats. For each splat, we test whether the condition from \Eq~\ref{eq:hessian-condition} is respected. If not, we normalize its eigenvalues to get $\lambda_1$ and $\lambda_2$, switch to log space and round them down and up. We also round the opacity down and up to the next/previous tenth. This gives us the 3D index of 8 precomputed entries that we retrieve from the dictionary by simple index linearization.

\paragraph{Resampling criterion} To test whether a splat with center $p$ and covariance $\Sigma$ satisfies the condition from~\Eq~\ref{eq:hessian-condition}, we verify that this criterion is satisfied with high probability when $y$ is drawn from the Gaussian distribution with mean $p$ and covariance matrix $\Sigma$.
More precisely, we randomly draw $512$ samples from $\mathcal{N}(p, \Sigma)$ and check whether at least $\tau = 95\%$ of the samples satisfy criterion~\Eq~\ref{eq:hessian-condition}. Note that in this model, $\varepsilon$ is a hyperparameter of our method; we perform a hyperparameter search over $\varepsilon \in \{0.001, 0.01, 0.1, 1.0\}$ to ensure that we resample sufficiently to avoid artifacts, while minimizing unnecessary resampling of splats. 

\paragraph{Interpolation and extrapolation} In the results reported below, we only use the closest neighbor among the 8 entries that we retrieve. We additionally tested linear interpolation, but found that it did not provide a meaningful improvement (see supplemental). %
We handle extreme eigenvalues, i.e. values beyond $10^4$, by clamping the lookup to the largest available training conditioning value, then extrapolating by rescaling the corresponding eigen-direction.%

\section{Results}

\subsection{Gradient rig} \label{sec:result-grad-rig}

We test our skinned deformation technique in a Web-based real-time Gaussian splatting renderer. As we limit our sparse weight representation to 2 non-zero weights per splats, we add for each splat 2 node indices, 2 weights and 2 gradients. We encode each of these new attributes on 8 bits, adding a total of 10 bytes per splat.

\paragraph{Quality} The quality of our deformation method is illustrated in \Fig~\ref{fig:deformed}, where we compare deformation results obtained with and without the elastic term $R_i^{\text{(elastic)}}$ introduced in our formulation. Across all examples, we consider a range of deformations, including translation and torsion. These results also show that our approach greatly reduces the visual artifacts caused by a regular skinned deformation.
We also compare our method with the Real-time Large-scale Deformation proposed by Gao et al~\shortcite{gao2024realtime}. We show in \Fig~\ref{fig:deformed_comparison}
that our method can reproduce deformation that are at least as large as their method. Furthermore, we do so solely in the space of Gaussian splats, without relying on a mesh proxy nor re-optimizing the Gaussians, which makes our approach readily available for other optimizers such as FastGS~\cite{ren2026fastgs}. \elieUpdate{Similarly to \citet{gao2024realtime}, our comparison focuses on qualitative measures per lack of ground truth for deformed objects (see also supplementary video).}

\paragraph{Performance} Since our deformation affects the render process, we measure its impact on rendering speed on an Apple M1 Max with 32 GB of memory on scenes ranging from 85K to 850K splats. Our study shows that the addition of our per-splat transform has no significant impact in this context. Indeed, the render process is largely fragment-bound; and a change of point of view have a much stronger effect on rendering speed than adding our deformation process, which is done once per vertex. In the end, the main runtime cost of our method is its impact on memory, since adding 10 bytes per splat might hinder the loading of larger scenes in GPU memory.

\subsection{Differentiable skinning weights}

\begin{figure}
    \centering
    \includegraphics[width=\linewidth]{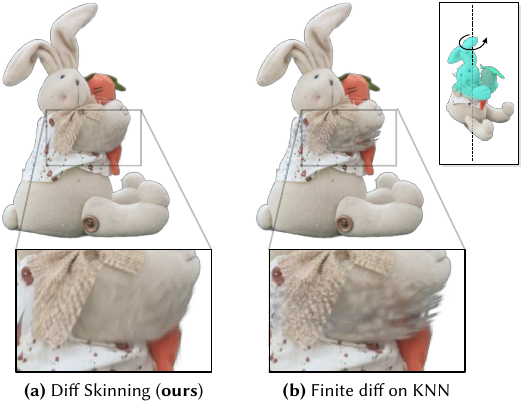}
    \caption{Our deformation procedure relies on an estimation of the gradients of skinning weights. We show that our differentiable skinning approach \textbf{(a)} described in \Sec~\ref{sec:autodiff-authoring} provides gradients that yield to deformation that exhibit less artifacts than an estimation based on finite differences on K nearest neighbors \textbf{(b)}. Top-right: the smooth selection mask and twist transform used for this example.}
    \label{fig:diff-skinning}
\end{figure}

\paragraph{Quality} We demonstrate that our automatic derivation of the gradient of skinning weights from the authoring gestures leads to more accurate results than estimating gradients afterwards. In \Fig~\ref{fig:diff-skinning}, we compare visually our approach with gradients estimated using Moving Least Squares on 16 nearest neighbors (KNN). In \Fig~\ref{fig:vs-knn}, we show through a quantitative analysis that this KNN estimation cannot match our differentiable skinning technique when the gradient varies quickly. This ablation confirms that the extra information that we harvest from user gestures when building an analytic definition of the weight field in \Sec~\ref{sec:diff-weights} can meaningfully improve the quality of gradients used for elastic deformation.

\paragraph{Performance} In \Tab~\ref{tab:overhead} we measure the runtime cost of computing gradients of skinning weights as part of every authoring manipulation that modifies them. With our auto-grad addition, weight edition is about 3 times slower, mostly because it computes 4 values (weight and spatial gradient) instead of one (weight only). It remains nonetheless interactive, and could be easily parallelize on GPU to better scale to large scenes. Furthermore, it is already about 10 times faster than the estimation using KNN. Note that this overhead is only paid once during the one-time setup that defines the skinning, not at deformation time (as reported in \Sec~\ref{sec:result-grad-rig}).

\begin{table}[ht]
    \centering
    \caption{Measure of the overhead introduced by our gradient computation during operations that modify skinning weights, in ms. Measured for a soft rectangle selection, on $N=32$ samples, CPU-only (JavaScript). We also report the KNN duration (16 neighbors per splat, using scipy's cKDTree).}
    \label{tab:overhead}
    \begin{tabular}{lrrrr}
        \toprule
        Scene & \#splats & w/o grads & w/ grads & KNN \\
        \midrule
        banana   & 56k   & $7.72 \pm 1.4$ & $19.9 \pm 1.7$  & $280 \pm 4$    \\
        car      & 160k  & $18.4 \pm 1.5$ & $47.2 \pm 9.9$  & $1080 \pm 9$   \\
        flower   & 270k  & $28.2 \pm 4.8$ & $83.9 \pm 6.1$  & $960 \pm 17$   \\
        fountain & 850k  & $80.9 \pm 2.8$ & $258 \pm 12$    & $2400 \pm 20$  \\
        tractor  & 2700k & $262 \pm 16$   & $827 \pm 42$    & $7700 \pm 300$ \\
        \bottomrule
\end{tabular}
\end{table}

\subsection{Resampling} \label{sec:result-diff-weights}

\begin{figure*}
    \centering
    \vspace{-0.2cm}
    \begin{minipage}[t]{0.49\textwidth}

        \centering
        \includegraphics[width=\linewidth, trim=0 60cm 28cm 1cm, clip]{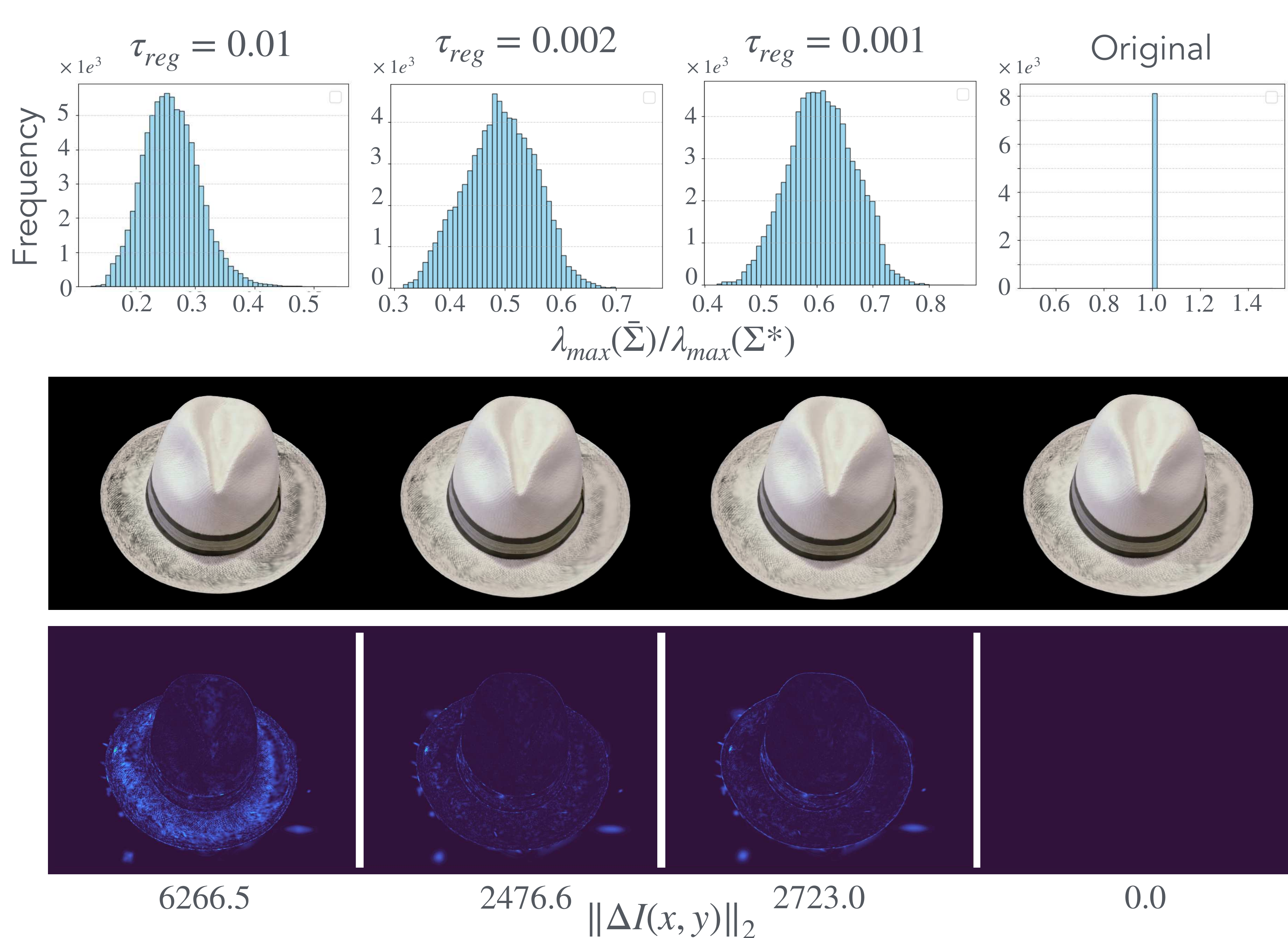}
        \caption{Impact of the regularization factor $\taureg$ used to balance the two terms of our loss when building the dictionary. Smaller $\taureg$ yield better preservation of the rest shape (middle and bottom rows), but higher $\taureg$ leads to a faster decrease of the largest eigenvalue, and thus to lower deformation error.}
        \label{fig:compare-taureg}

    \end{minipage}
    \hfill
    \begin{minipage}[t]{0.49\textwidth}
        \vspace{-6.2cm}
        \centering
        \includegraphics[width=\linewidth]{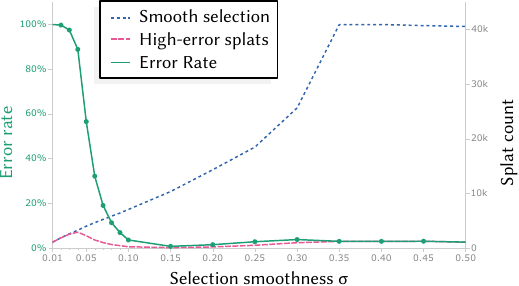}
        \caption{We compare our automatic differentiable skinning (\Sec~\ref{sec:diff-weights}) with Moving Least Square over 16 nearest neighbors (KNN). When the selection smoothness $\sigma$ is high enough, the weight field is smooth enough for the KNN method to correctly recover the gradient, but when $\sigma$ gets lower, the density of Gaussians is not enough to estimate steeper gradients. The error rate is measured among splats whose weight is neither 1 nor 0 and gradients is non null (which encompasses more splats when smoothness is higher). It is considered high when $|\Delta \grad w| > 1.0$ (our encoding allows gradients in range $(-4,4)$).}
        \label{fig:vs-knn}
    \end{minipage}
\end{figure*}

\begin{figure*}
    \centering
    \vspace{0.5cm}
    \includegraphics[width=0.95\linewidth]{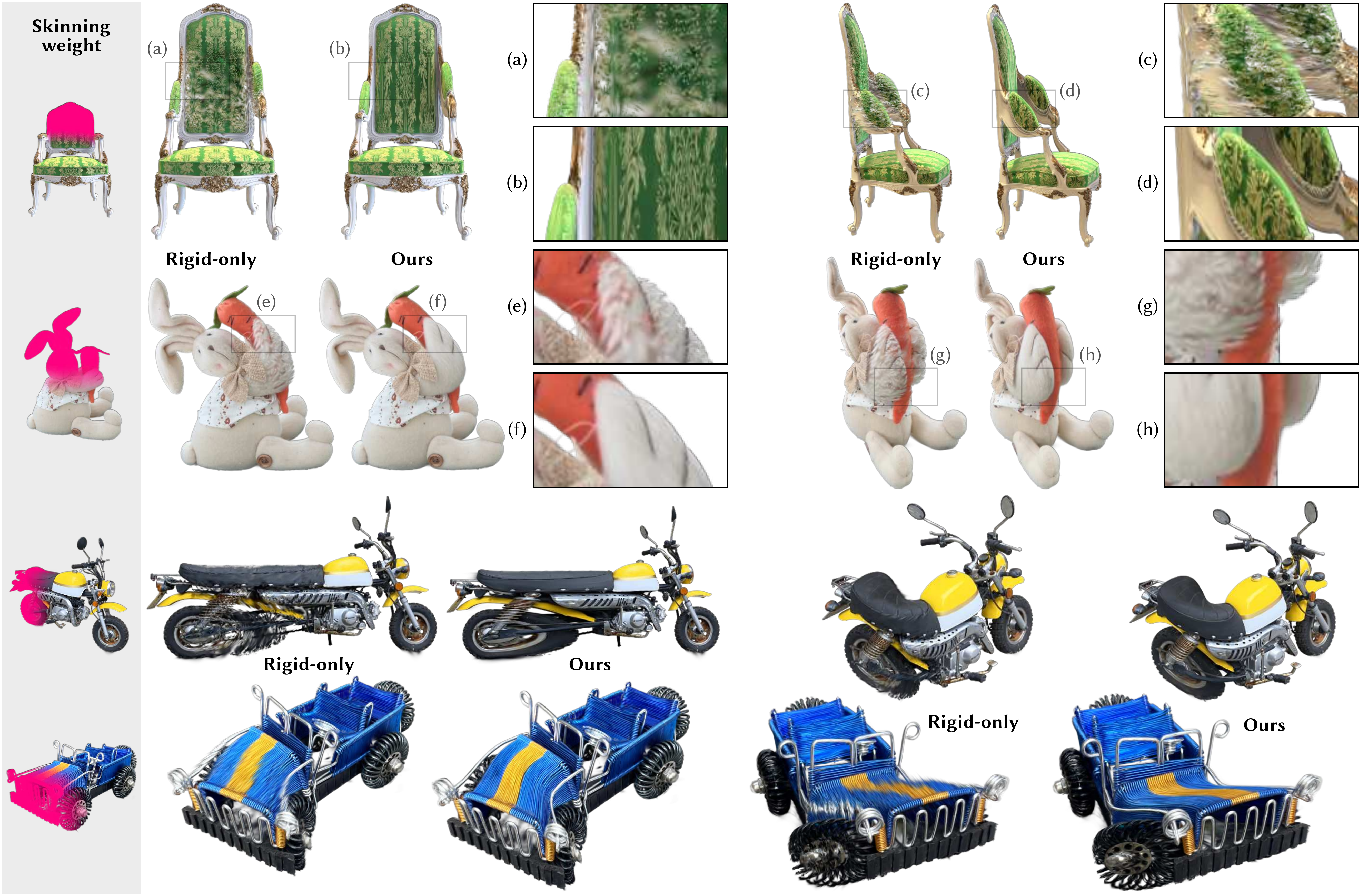}
    \caption{Results of our real-time deformation (\Sec~\ref{sec:grad-rig}) without any resampling. These examples feature large deformations, for which a rigid-only deformation introduces holes that our elastic term fixes. In all examples, weights and gradients are computed using smooth selections operations, as described in \Sec~\ref{sec:autodiff-authoring}.}
    \label{fig:deformed}
\end{figure*}

\begin{figure*}
    \centering
    \includegraphics[width=0.96\linewidth]{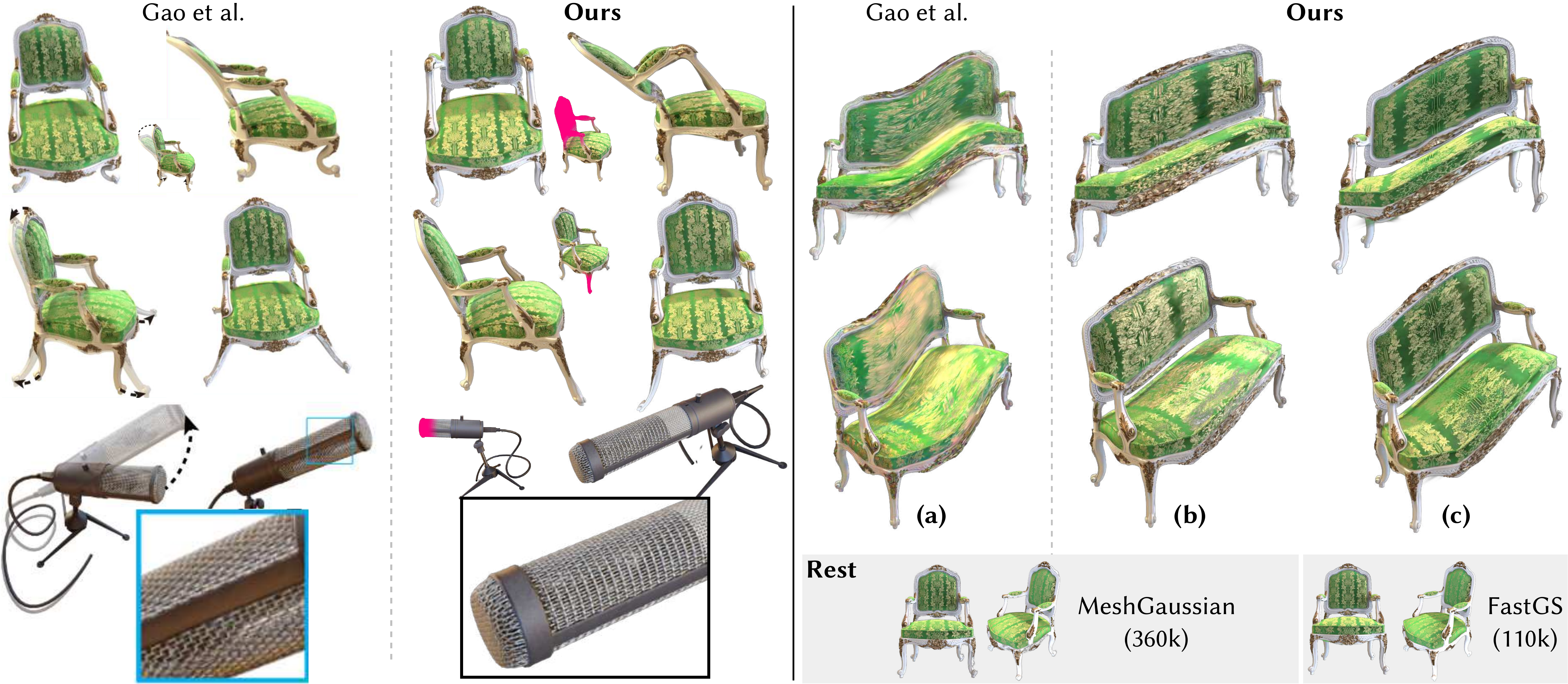}
    \caption{Comparison with \citet{gao2024realtime}. \textbf{Left:} images for Gao et al. are taken from their paper. We reproduce deformations of similar scale, showing that our method handles large-scale deformations without relying on a mesh proxy. \textbf{Right:} To compare similar deformations of the chair, \textbf{(a)} we run the code from Gao et al. and \textbf{(b)} we use our method to deform the splats resulting from their re-optimization of the asset. \textbf{(c)} Since our method does not require any specific splat optimization, we can apply it on the same chair optimized using FastGS~\shortcite{ren2026fastgs} (which trains in 2 min instead of 30 min and needs 60\% less Gaussians).}
    \label{fig:deformed_comparison}
\end{figure*}

\paragraph{Quality}
We analyze in \Fig~\ref{fig:compare-taureg} the impact of the regularization coefficient $\taureg$, that drives the trade-off between $\mathcal L_\text{rendering}$ and $\mathcal L_\text{conditioning}$ during the construction of our dictionary (\Eq~\eqref{eq:loss_resampling}). Overall, $\taureg$ must be picked as high as possible, within the limit of acceptable error in rest pose. Based on this experiment, we use $\tau_{reg}=0.002$, which reduces the diameter of the largest splat by approximately a factor of two on average, while keeping artifacts introduced by the resampling operation visually imperceptible. %

\begin{figure*}
    \centering
\includegraphics[width=0.95\linewidth, trim=0 50cm 0cm 0.7cm, clip]{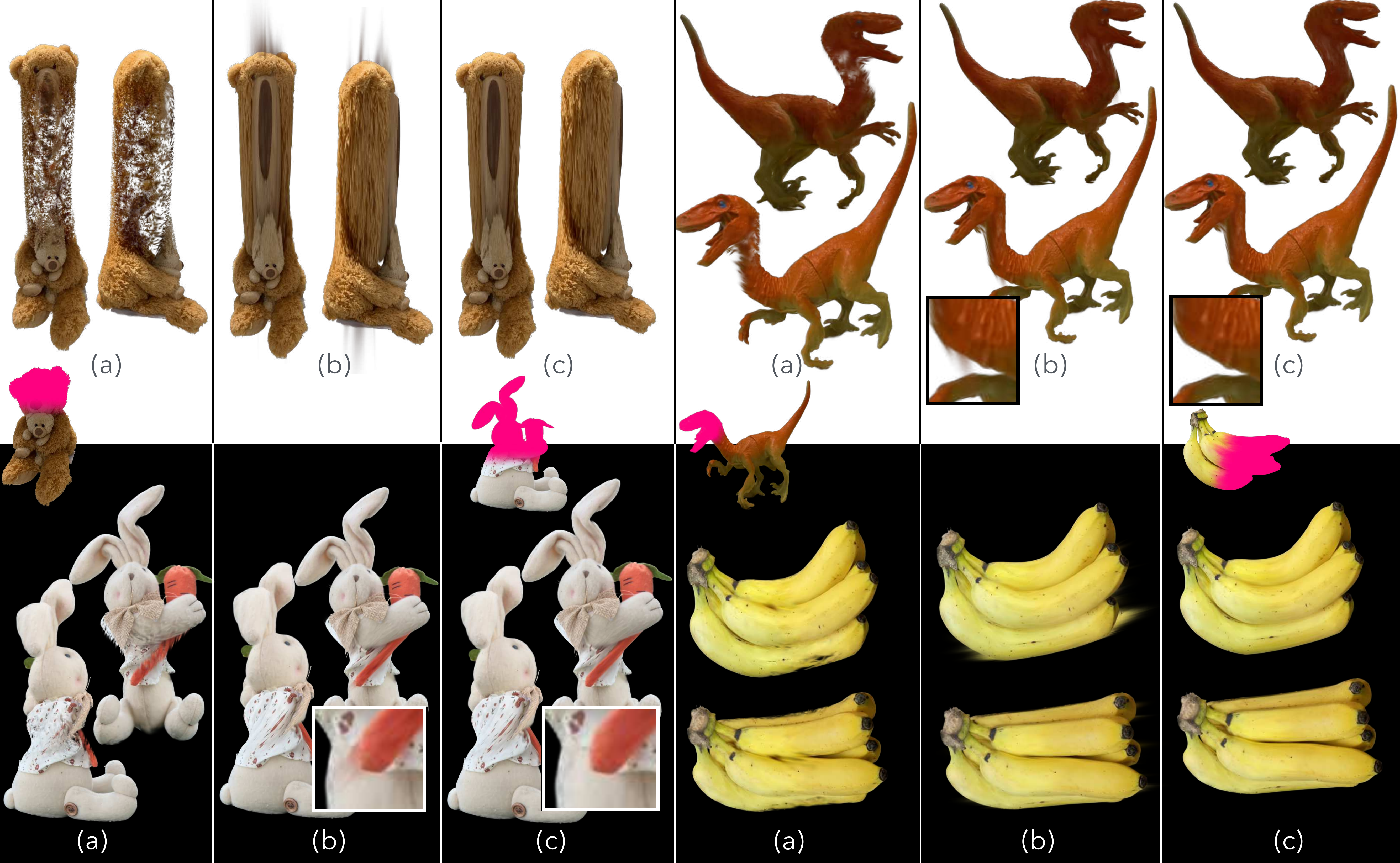}
    \caption{Four examples where the deformation field varies too rapidly for the affine approximation to remain valid. Naïve rigid deformation introduces holes (a). Our method resolves these holes, but some splats become overstretched due to strong deformation variations across their spatial extent (b). Our resampling procedure removes these artifacts (c). Adaptive resampling takes 82.2 s, 81.7 s, 21.8 s, and 9.4 s for the banana, teddy bear, rabbit, and dinosaur, respectively.}
    \label{fig:resampled_four}
\end{figure*}

We then evaluate in \Fig~\ref{fig:resampled_four} our resampling algorithm on several examples in which the deformation vector field varies rapidly; in particular, the splat weighting function is not smooth and has a large Lipschitz constant, which leads to visible artifacts. Across all examples, we observe that resampling significantly improves the visual quality of the deformed objects by preventing the splats to be overstretched. See supplemental for extra results and an analysis of the threshold hyperparameter $\varepsilon$.

\elieUpdate{
\paragraph{Performance} Since our resampling process is not mandatory for the user to get a real-time preview of the deformation (playing with $\eta$ gives satisfying results already), and since the resampling can be reused across deformations because it operates on the rest shape, we can afford to have a slower process. The adaptive resampling takes 10\,s to 1\,min 30\,s on the models in \Fig~\ref{fig:resampled_four}. This duration is largely dominated by the stochastic estimation of the adaptive criterion \Eq~\eqref{eq:hessian-condition} because uniformly applying the dictionary to the whole cloud takes about 1\,s for 160k splats (using numpy, CPU).
}

\section{Limitations and future work}

\paragraph{Gradient rig} Despite our use of a sparse encoding, explicitly storing the gradients of skinning weights introduces a memory overhead, which increases with the maximum number of rig nodes that influence a splat. If this would become an issue, we could improve the synergy between our gradient rig (\Sec~\ref{sec:grad-rig}) and differentiable skinning (\Sec~\ref{sec:diff-weights}) by storing only the symbolic representation of the weights (resulting from the composition of user selection gestures), and evaluating gradients on the fly. This would trade memory usage for compute power, but as we have seen the rendering process is not compute-bound.
We show results and derive equations only for the case of rigged deformation based on linear blend skinning. Our approach could be ported to other skinning models such as dual quaternion to enable more complex deformations.

\paragraph{Interpolation in the lookup table}
Our resampling procedure uses the closest entry from the precomputed dictionary, and we see in supplementary material that linear interpolation alone does not improve the results, but future work could explore other interpolation strategies. The interpolation scheme for opacity could be more consistent with the rendering formula, where rendering
two Gaussians is not equivalent to summing their opacities. As for interpolating the rotations, a promising direction would be to use the Wasserstein barycenter. In fact, the barycenter of a Gaussian distribution is itself Gaussian~\citep{agueh2011barycenters}. However, there is no closed-form expression for it, and computing it would require solving the optimization problem defined by Equation~(6.2) in~\citet{agueh2011barycenters} during resampling, which is too costly in practice. A useful direction for future work would therefore be to find a cheap surrogate for this procedure.

\paragraph{Resampling for a whole deformation space}
Our resampling strategy consists in resampling the rest shape, rather than the deformed shape. The strength of this approach is that we can reuse the resampling computed for a given transform on other neighboring deformations, and for in-between with the rest shape, where the error is overall lower. We could imagine generalizing this further by relying on the maximum amplitude of plausible deformations of each rig node (e.g., the elbow node is only rotating around a given axis and within a given range) to resample the Gaussian splat cloud for a whole class of transforms, ahead of time. \elieUpdate{Meanwhile, we could improve its runtime performance by accelerating the estimation of erroneous Gaussian splats and resampling on GPU, since our dictionary-based approach is very parallel-friendly. To further reduce the introduction of new splats, we could also have the resampling process tune a per-splat elastic factor $\eta_i$, to selectively reduce the elongation only for splats that cause artifacts.}

\elieUpdate{\paragraph{Conclusion} We propose a method to interactively deform Gaussian splats without requiring a mesh proxy, nor a specific re-optimization or view-specific projections. We also provide a rest-space resampling to further improve the quality of deformed shapes.}

\bibliographystyle{ACM-Reference-Format}
\bibliography{biblio}

\newpage

\clearpage

\section*{Supplementary Material}

\section{Justification of the affine approximation}

Given a diffeomorphism $F$ and a probability measure $\rho$, the density associated with $\gamma = F_\# \rho$ (where $F_\# \rho$ denotes the pushforward distribution of $X \sim \rho$ under $F$) is given by
$$\gamma(x) = \frac{\rho(F^{-1}(x))}{\det(J_F(F^{-1}(x)))},$$
where $J_F$ is the Jacobian of $F$. When $\rho$ is a mixture of Gaussians, $\rho = \sum_{i=1}^K r_i \mathcal{N}(p_i, \Sigma_i),$
the resulting probability measure is generally not a mixture of Gaussians. In particular this is not the case %
for the deformation field we consider due to the non-linearity of the skinning weight functions $w_j$. However, when $F$ is affine, i.e., $F(x) = Gx + h$, the resulting $\gamma$ is also a mixture of Gaussians with the following expression:
$$\gamma = \sum_{i=1}^K r_i \mathcal{N}(G p_i + h,\, G \Sigma_i G^T).$$
For this reason, and similarly to~\citet{xie2024physgaussian}, we propose to perform local affine approximations of the vector field on each Gaussian in order to remain in the space of Gaussian mixtures. Each particle (Gaussian) is deformed by a local affine transformation given by the first-order Taylor expansion of $F$ at the center of the Gaussian $p$. More precisely, we approximate $F$ as
$$F(y) \approx F(p) + J_F(p)(y - p),$$
for points $y$ belonging to the Gaussian. Consequently, each Gaussian $\mathcal{N}(p, \Sigma)$ is mapped to
$$\mathcal{N}\big(F(p),\, J_F(p)\,\Sigma\,J_F(p)^T\big).$$

\section{Jacobian of the deformation field}

In \Sec~3.1 of the main paper, we use the Jacobian $R$ of the deformation field $F(p)$ and decompose it into $R^{(\text{rigid})} + R^{(\text{elastic})}$. We give here details about the derivation. In what follows, all gradients $\grad$ are with respect to the spatial coordinates of $p$ and $J_F = \partial F / \partial p$ represents the Jacobian of $F$ with respect to $p$.

We recall that
\begin{equation*}
    F(p)
    =
    \frac{1}{W(p)}
    \sum_{j=0}^{M-1} w_j(p) T'_j p,
    \qquad
    W(p)=\sum_{j=0}^{M-1} w_j(p).
\end{equation*}
Let us introduce the normalized weights
\begin{equation*}
    \alpha_j(p)=\frac{w_j(p)}{W(p)}.
\end{equation*}
Then
\begin{equation*}
    F(p)
    =
    \sum_{j=0}^{M-1}\alpha_j(p)T'_j p.
\end{equation*}
We also define
\begin{equation*}
    A_j = \mathrm{mat3x3}(T'_j),
    \qquad
    q_j(p)=T'_j p.
\end{equation*}
Thus,
\begin{equation*}
    F(p)=\sum_{j=0}^{M-1}\alpha_j(p)q_j(p).
\end{equation*}
Since $q_j$ is affine in $p$, we have
\begin{equation*}
    J_{q_j}=A_j.
\end{equation*}
Applying the product rule gives
\begin{align*}
    J_F
    &=
    \sum_{j=0}^{M-1}
    J_{\alpha_j q_j} \\
    &=
    \sum_{j=0}^{M-1}
    \left(
        \alpha_j A_j
        +
        q_j \otimes \nabla \alpha_j
    \right).
\end{align*}
The first term corresponds to the rigid part:
\begin{equation*}
    R^{(\mathrm{rigid})}
    =
    \sum_{j=0}^{M-1}\alpha_j A_j
    =
    \mathrm{mat3x3}
    \left(
        \sum_{j=0}^{M-1}\alpha_j T'_j
    \right).
\end{equation*}
For the second term, we compute
\begin{align*}
    \nabla \alpha_j
    &=
    \nabla\left(\frac{w_j}{W}\right) \\
    &=
    \frac{\nabla w_j}{W}
    -
    \frac{w_j}{W^2}\nabla W.
\end{align*}
Using
\begin{equation*}
    \nabla W
    =
    \sum_{j'=0}^{M-1}\nabla w_{j'},
\end{equation*}
we define
\begin{equation*}
    \vec G_j
    =
    \nabla\alpha_j
    =
    \frac{\nabla w_j}{W}
    -
    \frac{w_j}{W^2}
    \sum_{j'=0}^{M-1}\nabla w_{j'}.
\end{equation*}
Therefore,
\begin{equation*}
    J_F
    =
    R^{(\mathrm{rigid})}
    +
    \sum_{j=0}^{M-1} q_j \otimes \vec G_j.
\end{equation*}
Finally, since $q_j=T'_j p$, we obtain
\begin{align*}
    J_F
    &=
    R^{(\mathrm{rigid})}
    +
    R^{(\mathrm{elastic})}, \\
    R^{(\mathrm{rigid})}
    &=
    \sum_{j=0}^{M-1}\alpha_j \mathrm{mat3x3}(T'_j), \\
    R^{(\mathrm{elastic})}
    &=
    \sum_{j=0}^{M-1}(T'_j p)\otimes \vec G_j.
\end{align*}

\section{Hessian of the $k$-th coordinate of the deformation field}
 \label{ap:field-hessian}

We now derive the Hessian of the $k$-th coordinate of the deformation field
using the same notation as in the Jacobian derivation. We recall that
\begin{equation*}
    F(p)=\sum_{j=0}^{M-1}\alpha_j(p)q_j(p),
    \qquad
    q_j(p)=T'_j p,
    \qquad
    A_j=\mathrm{mat3x3}(T'_j),
\end{equation*}
with
\begin{equation*}
    \alpha_j(p)=\frac{w_j(p)}{W(p)},
    \qquad
    W(p)=\sum_{j=0}^{M-1}w_j(p).
\end{equation*}
For the $k$-th coordinate,
\begin{equation*}
    F_k(p)
    =
    \sum_{j=0}^{M-1}\alpha_j(p)q_{j,k}(p).
\end{equation*}
Since $q_j$ is affine in $p$, we have
\begin{equation*}
    \nabla q_{j,k}=A_{j,k:},
    \qquad
    \nabla^2 q_{j,k}=0,
\end{equation*}
where $A_{j,k:}$ denotes the $k$-th row of $A_j$, interpreted as a 3D vector.

From the Jacobian derivation, we have
\begin{equation*}
    \vec G_j
    =
    \nabla \alpha_j
    =
    \frac{\nabla w_j}{W}
    -
    \frac{w_j}{W^2}\nabla W.
\end{equation*}
Applying the product rule twice gives
\begin{align*}
    H_{F_k}
    =
    \nabla^2 F_k
    &=
    \sum_{j=0}^{M-1}
    \nabla^2\left(\alpha_j q_{j,k}\right) \\
    &=
    \sum_{j=0}^{M-1}
    \left(
        q_{j,k}\nabla^2\alpha_j
        +
        \nabla q_{j,k}\otimes\nabla\alpha_j
        +
        \nabla\alpha_j\otimes\nabla q_{j,k}
        +
        \alpha_j\nabla^2 q_{j,k}
    \right).
\end{align*}
Using $\nabla^2 q_{j,k}=0$ and $\nabla\alpha_j=\vec G_j$, we obtain
\begin{equation*}
    H_{F_k}
    =
    \sum_{j=0}^{M-1}
    \left(
        q_{j,k}\nabla\vec G_j
        +
        A_{j,k:}\otimes\vec G_j
        +
        \vec G_j\otimes A_{j,k:}
    \right).
\end{equation*}
There remains to expand $\nabla\vec G_j$. Starting from
\begin{equation*}
    \vec G_j
    =
    \frac{1}{W}\nabla w_j
    -
    \frac{w_j}{W^2}\nabla W,
\end{equation*}
we differentiate the first term:
\begin{equation*}
    \nabla\left(\frac{1}{W}\nabla w_j\right)
    =
    \frac{1}{W}\nabla^2 w_j
    -
    \frac{1}{W^2}\nabla w_j\otimes\nabla W.
\end{equation*}
For the second term, we have
\begin{align*}
    \nabla\left(\frac{w_j}{W^2}\nabla W\right)
    &=
    \nabla\left(\frac{w_j}{W^2}\right)\otimes\nabla W
    +
    \frac{w_j}{W^2}\nabla^2 W \\
    &=
    \left(
        \frac{1}{W^2}\nabla w_j
        -
        \frac{2w_j}{W^3}\nabla W
    \right)\otimes\nabla W
    +
    \frac{w_j}{W^2}\nabla^2 W.
\end{align*}
Therefore, we define $\mathbf K_j$ as the gradient of $\vec G_j$:
\begin{align*}
    \mathbf K_j
    &:=
    \nabla\vec G_j \\
    &=
    \frac{1}{W}\nabla^2 w_j
    -
    \frac{1}{W^2}\nabla w_j\otimes\nabla W \\
    &\quad
    -
    \left(
        \frac{1}{W^2}\nabla w_j
        -
        \frac{2w_j}{W^3}\nabla W
    \right)\otimes\nabla W
    -
    \frac{w_j}{W^2}\nabla^2 W \\
    &=
    \frac{1}{W}\nabla^2 w_j
    -
    \frac{2}{W^2}\nabla w_j\otimes\nabla W
    -
    \frac{w_j}{W^2}\nabla^2 W
    +
    \frac{2w_j}{W^3}\nabla W\otimes\nabla W \\
    &=
    \frac{1}{W}\nabla^2 w_j
    -
    \frac{w_j}{W^2}\nabla^2 W
    -
    \frac{2}{W}\left(
        \frac{1}{W}\nabla w_j
        -
        \frac{w_j}{W^2}\nabla W
    \right)\otimes\nabla W \\
    &=
    \frac{1}{W}\nabla^2 w_j
    -
    \frac{w_j}{W^2}\nabla^2 W
    -
    \frac{2}{W}\vec G_j\otimes\nabla W.
\end{align*}
Finally, using $q_{j,k}=(T'_j p)_k$, we get
\begin{equation*}
    H_{F_k}
    =
    \sum_{j=0}^{M-1}
    \left(
        (T'_j p)_k\mathbf K_j
        +
        A_{j,k:}\otimes\vec G_j
        +
        \vec G_j\otimes A_{j,k:}
    \right).
\end{equation*}
Thus, the Hessian of the $k$-th coordinate is
\begin{align*}
    &H_{F_k}
    =
    \sum_{j=0}^{M-1}
    \left(
        (T'_j p)_k\mathbf K_j
        +
        A_{j,k:}\otimes\vec G_j
        +
        \vec G_j\otimes A_{j,k:}
    \right), \\
    &\text{with} \quad
    \vec G_j
    =
    \frac{\nabla w_j}{W}
    -
    \frac{w_j}{W^2}\nabla W, \\
    &\text{and} \quad
    \mathbf K_j
    =
    \frac{1}{W}\nabla^2 w_j
    -
    \frac{w_j}{W^2}\nabla^2 W
    -
    \frac{2}{W}\vec G_j\otimes\nabla W.
\end{align*}
Note that when the weights sum to one, so that $W(p)=1$ is constant, $\mathbf K_j$ reduces to $\nabla^2 w_j$ and $G_j$ is equal to $\nabla w_j$.
Moreover,
\begin{equation*}
    \nabla W=\sum_{j'=0}^{M-1}\nabla w_{j'},
    \qquad
    \nabla^2 W=\sum_{j'=0}^{M-1}\nabla^2 w_{j'}.
\end{equation*}
The entries of $H_{F_k}$ are
\begin{equation*}
    (H_{F_k})_{ab}
    =
    \frac{\partial^2 F_k}{\partial p_a\partial p_b}.
\end{equation*}

In this work, our first-order approximation is valid as long as the second-order term in the Taylor expansion of $F$ remains small. Using a second-order Taylor expansion around $p$, we have
\begin{equation*}
F(y)
\approx
F(p)
+
J_F(p)(y - p)
+
\frac{1}{2}
\begin{pmatrix}
    (y-p)^T H_{F_1}(p)(y-p) \\
    \vdots \\
    (y-p)^T H_{F_d}(p)(y-p)
\end{pmatrix}.
\end{equation*}
Therefore, the validity of the first-order approximation depends on the magnitude of the quadratic terms
\begin{equation*}
    (y-p)^T H_{F_k}(p)(y-p),
    \quad k = 1, \dots, d.
\end{equation*}

\section{Maximum number of rig node per splat}

\begin{figure}
    \centering
    \includegraphics[width=\linewidth]{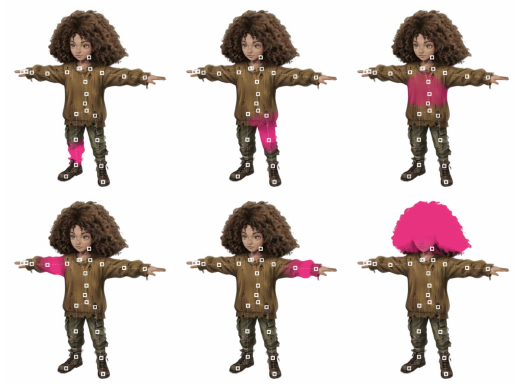}
    \caption{\elieUpdate{We test our deformation technique with the skinning weights produced by RigAnything~\cite{liu2025riganything}. In this example, 8.5\% of Gaussian splats are significantly influenced by more than 2 rig nodes.}}
    \label{fig:riganything}
\end{figure}

\elieUpdate{Our experiments are limited to a maximum of 2 rig nodes influencing a given Gaussian splat in order to limit the number of attributes to add for each splats. When there is more than 2 rig nodes influencing a Gaussian splat, two option are possible: \textbf{(a)} the easiest one is to add another pair of attribute for each splats (rig node index + weight) and all the formulas generalize well or (b) if the memory overhead of these extra attributes is too costly for the application case, one can introduce new mock rig nodes that track weighted barycenters of the true rig nodes. In \Fig~\ref{fig:riganything} we measure on a real scenario (the output of RigAnything~\cite{liu2025riganything}) the number of splats that are influenced by more than 2 rig nodes. Limiting to 2 rig nodes is a problem for 8.5\% of Gaussian splats, which are typically located on the hands and feet, were they cause some splats to be detached from one of their nodes. Increasing to 3 rig nodes addresses the issue.}

\section{Interpolation in dictionary lookup}

In \Sec~3.3 of the main paper, we precompute a dictionary of resampled Gaussians. After a deformation, we look this up to selectively resample all Gaussians of a shape. We show in \Fig~\ref{fig:resampling_interpolation_vs_nearest} that with the resolution of our dictionary, linear interpolation does not significantly improve results over nearest neighbor interpolation (which is less costly).

\section{Range of eigenvalues}

\begin{figure}[h]
    \centering
    \includegraphics[width=\linewidth, trim=0cm 87cm 50cm 3cm, clip]{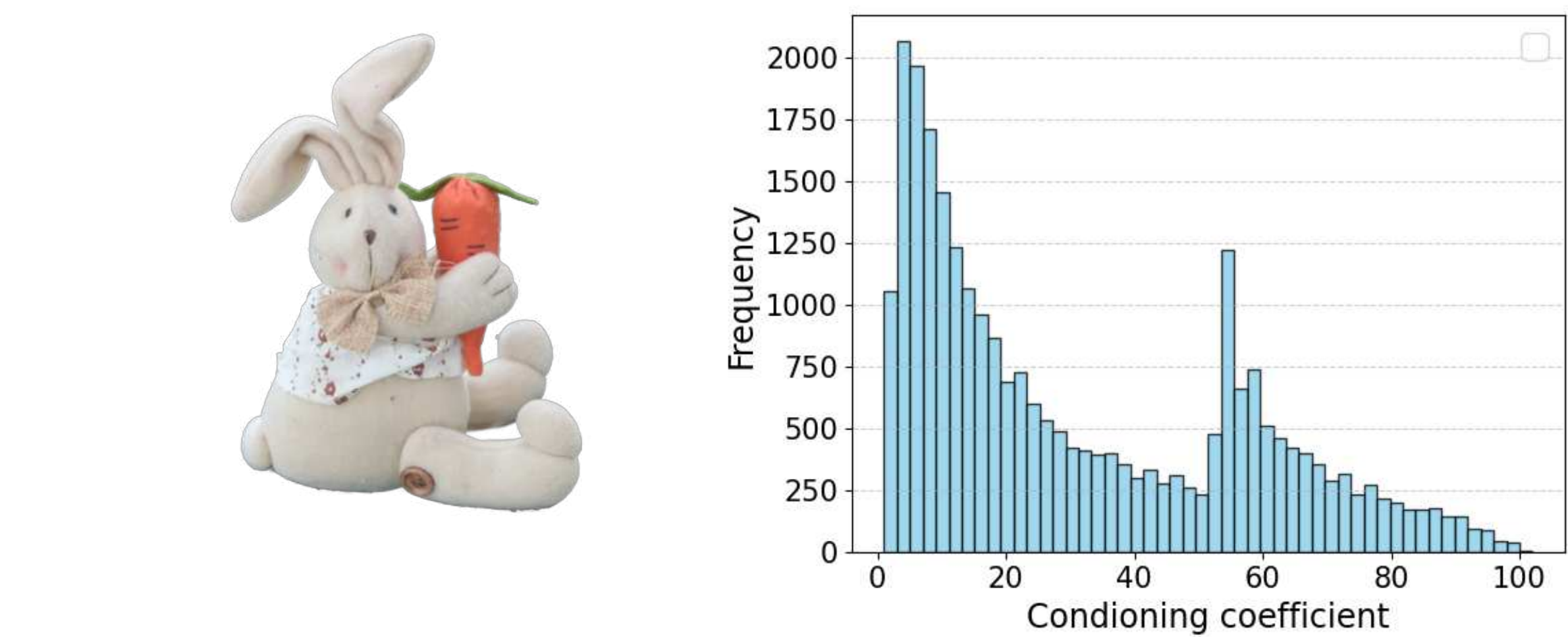}
    \includegraphics[width=\linewidth, trim=47cm 87cm 4cm 3cm, clip]{figures/resampling/cond_coeff.pdf}
    \caption{ Distribution of conditioning coefficients ($\lambda_{\max}/\lambda_{\min}$) for all Gaussian splats of the rabbit and the wire car model.}
    \label{fig:conditionning-distrib}
\end{figure}

To determine the discretization scale used for our precomputed dictionary, we study the range of eigenvalues different splat scenes. More precisely, we study the ratio $\lambda_{\max} / \lambda_{\min}$, since Gaussians are normalized before looking up the dictionary. \Fig~\ref{fig:conditionning-distrib} shows that it is typically between $1$ and $10^4$, but can become very large (up to $10^{12}$ in our worst cases). We use this analysis to establish the range of our precomputed dictionary.

\section{Extra result for our gradient-based deformation}

In \Fig~\ref{fig:deformed_other}, we present additional deformation examples similar to those shown in \citet{gao2024realtime}. These results demonstrate that our deformation procedure can readily reproduce deformations of comparable scale and nature to those of \citet{gao2024realtime}.

\section{Estimating gradients using finite differences on neighbor graph} \label{sec:skin-baseline}

\subsection{Method}

When we evaluate our differentiable skinning weight definition in \Sec~4.2 of the main paper, we compare it with an \textit{a posteriori} estimation that we perform using finite differences. If we already know the value $w_{i,j}$ of the skinning weight field $w_j$ at all positions $(p_i)_{0 \leq i < N}$, we can estimate the gradients using Moving Least Squares (like point set surfaces do to define their normal \cite{Alexa01,Guennebaud07}). For a position $p$, the gradient of $w_j$ is estimated by minimizing a local least square error:
\begin{equation}
    \grad w_j(p) = \argmin_{\vec g} \sum_{i=0}^{N-1} \alpha_i \Big((w_j(p) - w_j(p_i)) - \vec g \cdot (p - p_i)\Big)
\end{equation}
where influence weights $\alpha_i$ indicate how much $p_i$ is a neighbor of $p$. Typically, only a small number of $\alpha_i$ are non null, and they live in the vicinity of $p$ so we use a KD-Tree to look for the 16 nearest neighbors of $p$ and only account for these neighbors in the minimization.

Non-null weights $\alpha_i$ are computed as $1 / d_i^2$, where the distance $d_i$ measures how close we are to the splat $i$. We use the Euclidean distance between the Gaussian centers $p$ and $p_i$:
\begin{equation}
    d_i = \|p - p_i\|^2
\end{equation}

\subsection{Results}

In \Fig~7 of the main paper, we report for various values of feather (i.e., selection smoothness $\sigma$) the error between skinning gradients obtained using our differentiable skinning and a reconstruction of gradients from weights only using KNN. This experiment can be seen as an ablation: could we use our gradient-based deformation technique without differentiable skinning? For larger values of feather we can, but we have to be careful when splat density gets smaller wrt. gradient variation. Note that here we consider the result of our differentiable skinning to be the ground truth to compare KNN against because it has access to more information (namely, an analytic function that describes the weight field).

In the folder \texttt{data/ours-vs-knn} of this supplementary data, we share extra detailed about the comparison reported in \Fig~7 of the main paper. For each value of feather, we plot an histogram of the error. Feather shape was set to "smoothstep" (as opposed to "linear"). The file \texttt{stats.txt} contains the raw data of \Fig~7 and \texttt{chair\_selection.png} shows the scene for which we tested, with the skinning weight highlighted in orange.

\section{Comparison with Gao et al.}

\newcommand\fname[1]{\textbf{\texttt{#1}}}

In \Fig~10 of the main paper we present a comparison with results from \citet{gao2024realtime}. Since the figure only shows selected viewpoints, we also share in folder \texttt{data/ours-vs-gao} the PLY and OBJ files that were used. Note that for the sake of concision we only use the first letter of each filename (e.g., \fname{x} for \texttt{x\_foo.ext}).

Files \fname{a}, \fname{b} and \fname{c} correspond to the \textbf{(a)}, \textbf{(b)} and \textbf{(c)} columns of the right-hand side of \Fig~10. In particular, files \fname{a} (Gao et al.) and \fname{b} (ours) are both deformed versions of the same rest input, which is given in file \fname{d} (optimized by Gao et al.). In the case of file \fname{a} -- which is deformed by the method of Gao et al. -- only  the rest input from file \fname{d} could be used, whereas in file \fname{c} we show an example of our method applied on a rest input provided by a different optimizer (namely FastGS~\cite{ren2026fastgs}), provided in file \fname{e}.

This ability to adapt to any new Gaussian splat optimizer is a notable advantage of our method over the one proposed by Gao et al. Another benefit is that we do not need a mesh proxy; we share in file \fname{f} the mesh extracted from the rest splats and in file \fname{g} the deformed mesh, provided back to the pipeline of Gao et al. to evaluate the deformed splats. Deformation was done in Blender using the proportional editing tool, which is very close to our model for defining skinning weights but not perfectly identical, which is why we did not try to reproduce the exact same deformation.

\section{Additional results for resampling}

\paragraph{Impact of resampling.}
We validate in \Fig~\ref{fig:compare-rest} that applying our resampling procedure does not significantly change the appearance of the shape. For this experiment, we resample all splats and replace each of them by five smaller splats, without using our resampling criterion, in order to isolate the effect of the resampling process itself on the rendered object. Although the rendered views are not perfectly preserved, the error remains low and is well distributed over the shape.

Instead of selecting the closest splat in the lookup table, we also experimented with linearly interpolating between the four lookup-table entries corresponding to the splat to be resampled. We compare both strategies in \Fig~\ref{fig:resampling_interpolation_vs_nearest}. Linear interpolation does not appear to improve the results.

\begin{figure*}
    \centering
        \includegraphics[width=1.0\linewidth, trim=0 58cm 0cm 2, clip]{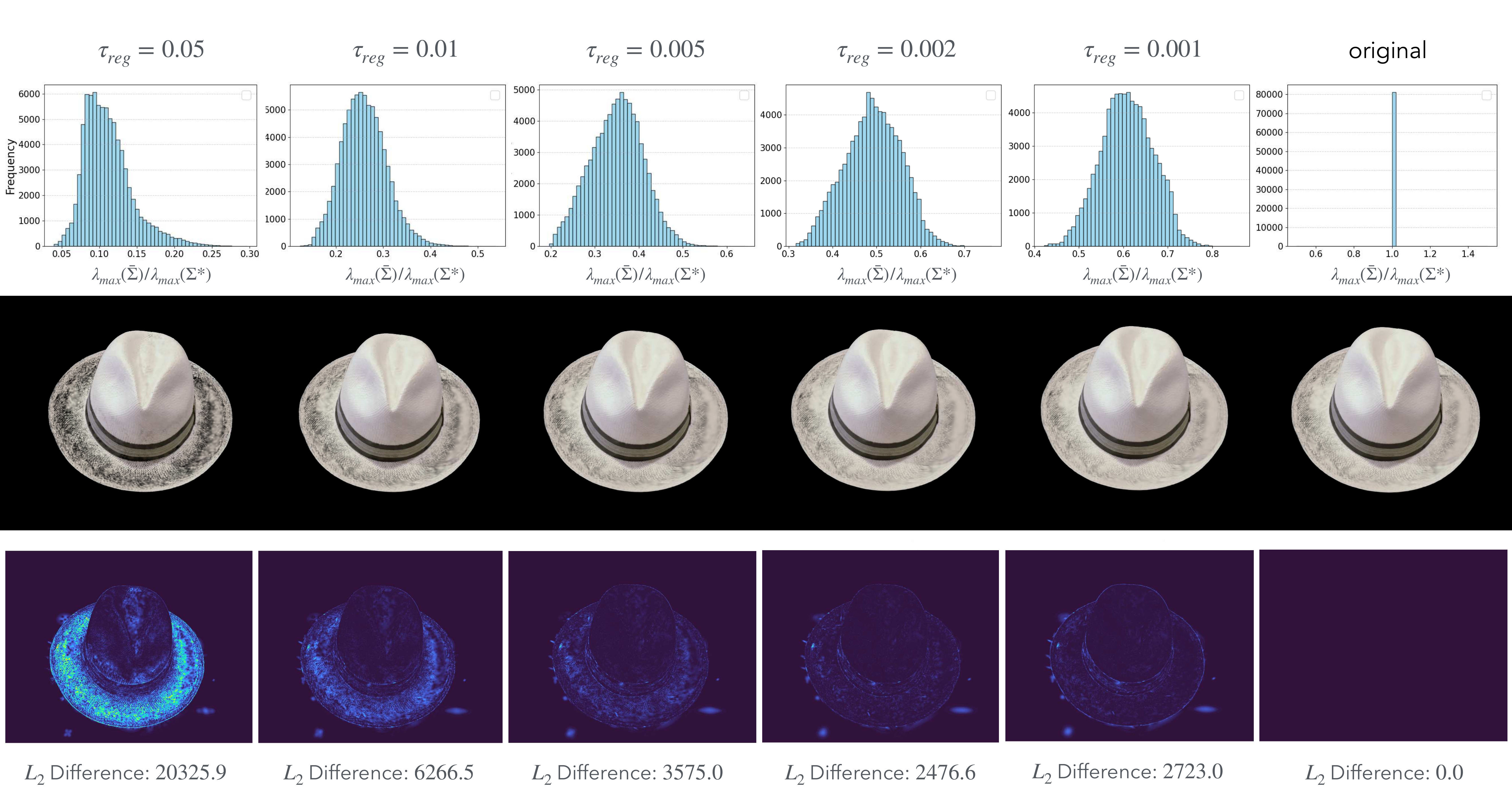}
        \caption{Impact of the regularization factor $\taureg$ used to balance the two terms of our loss when building the dictionary. Smaller $\taureg$ yield better preservation of the rest shape (middle and bottom rows), but higher $\taureg$ leads to a faster decrease of the largest eigenvalue, and thus to lower deformation error.}
        \label{fig:compare-taureg_annex}
\end{figure*}

\begin{figure*}
    \centering \includegraphics[width=1.0\linewidth, trim=0 73cm 0cm 0, clip]{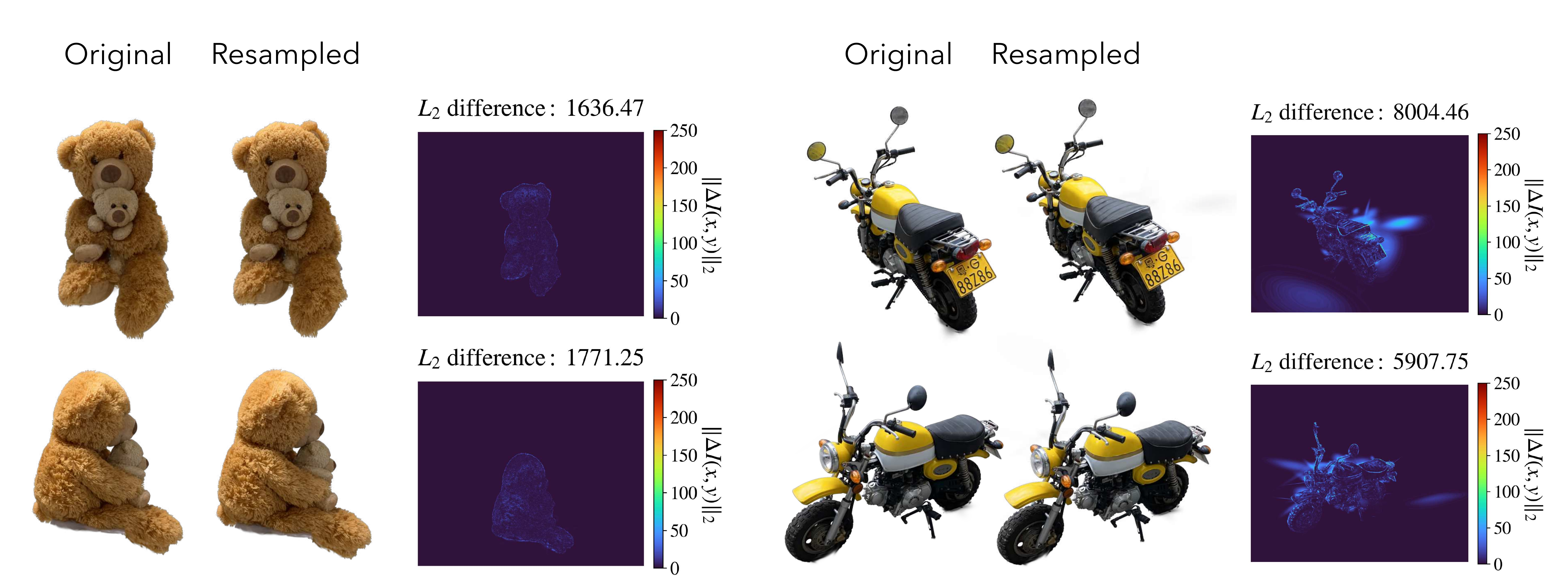}
    \caption{Comparison of images rendered from the original cloud of Gaussian splats (left) and from our resampled cloud (right). This validates that for the rest position, our resampling does not introduce significant error. The error map on the right shows that the error is well distributed over the shape (i.e., it does not introduce local artifacts).}
    \label{fig:compare-rest}
\end{figure*}

\begin{figure*}
    \centering
    \includegraphics[width=.9\linewidth]{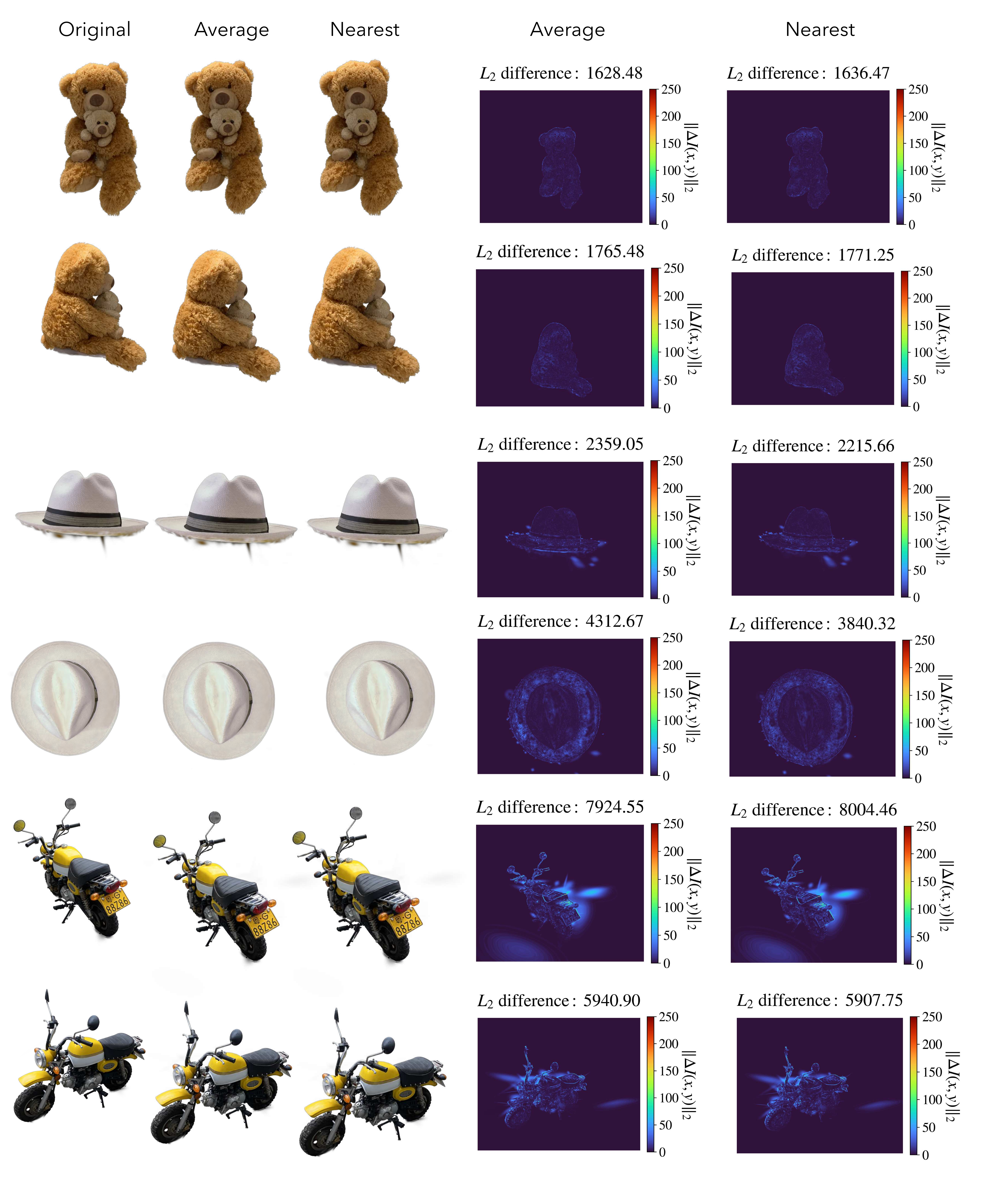}
    \caption{In our 2D look-up table, instead of selecting the nearest case for the splat to be resampled, we perform interpolation using the four neighboring cases and apply linear interpolation. The weights are defined as the inverse of the Euclidean distance between the eigenvalues of the target Gaussian and those of the corresponding cases in the table. This figure compares the nearest-neighbor and interpolation approaches. Both methods produce similar results, and neither method is consistently superior to the other.}
    \label{fig:resampling_interpolation_vs_nearest}
\end{figure*}

\begin{figure*}
    \centering
    \includegraphics[width=.95\linewidth]{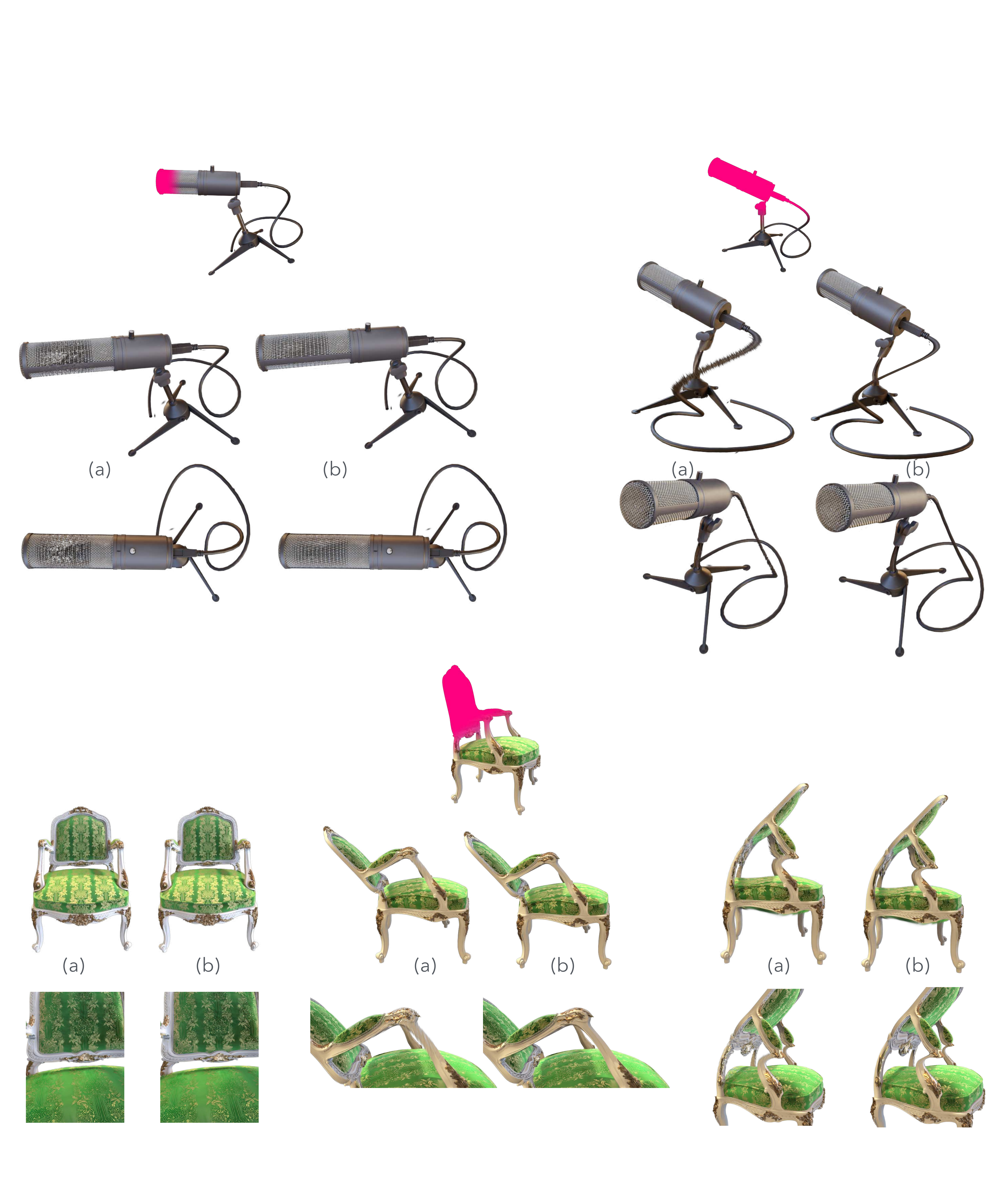}
    \caption{Other object deformations similar to~\citet{gao2024realtime}). The naïve rigid deformation creates holes in the Gaussian splats (a). Our method removes these artifacts and produces the intended deformation (b).}
    \label{fig:deformed_other}
\end{figure*}

\paragraph{Choice of $\tau_{\mathrm{reg}}$.}
To construct our lookup table, we need to choose a suitable value for the hyperparameter $\tau_{\mathrm{reg}}$. This parameter must both ensure an accurate reconstruction of the original splat and reduce the largest eigenvalue when replacing it with five smaller splats. In \Fig~\ref{fig:compare-taureg_annex}, we show an extended version of \Fig~7 from the main paper, which justifies our choice of $\tau_{\mathrm{reg}}=0.002$ as a good compromise.

\begin{figure*}
    \centering
    \includegraphics[width=.9\linewidth]{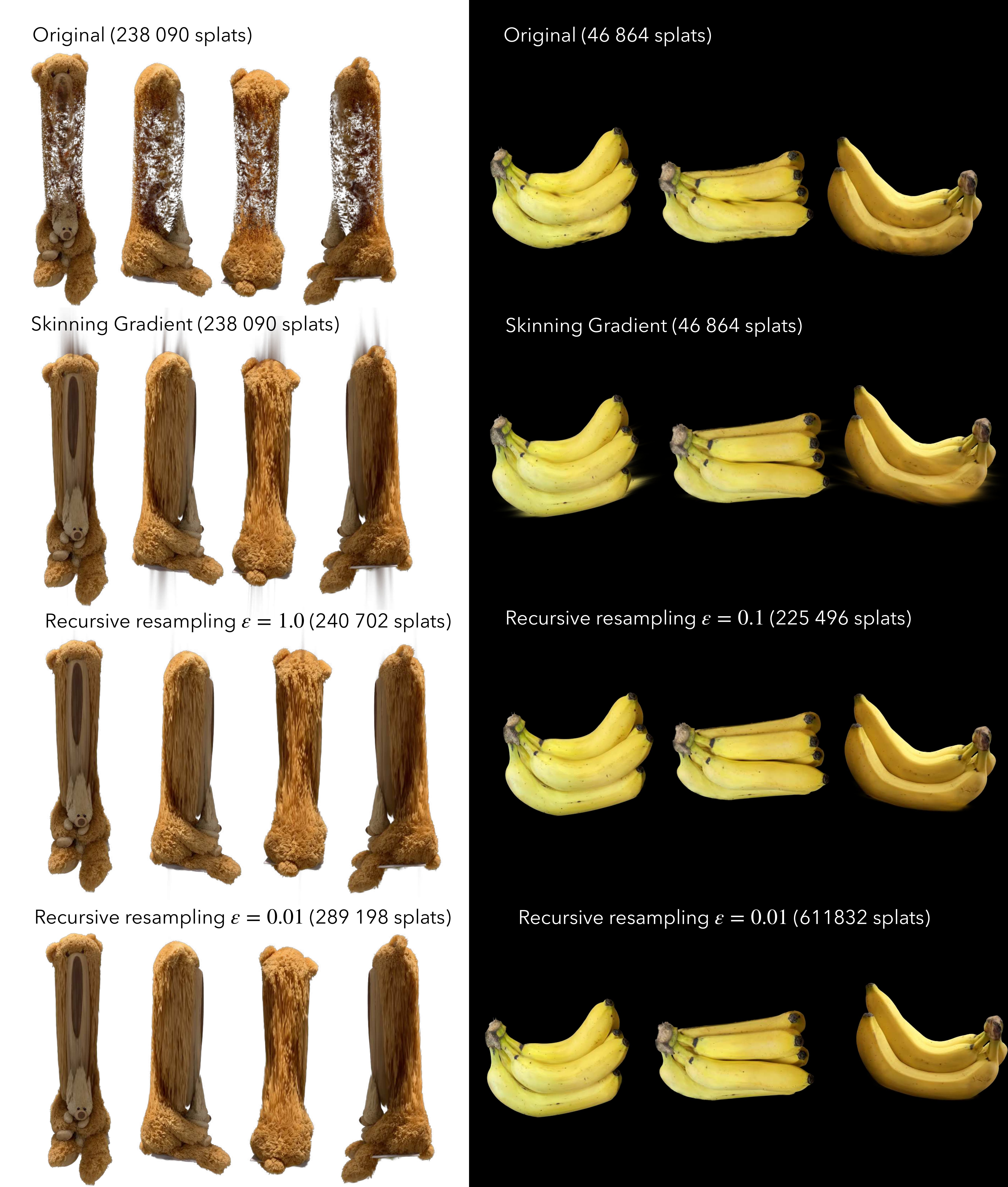}
    \caption{The threshold $\varepsilon$ quantifies when the second-order term is sufficiently small to avoid resampling a splat. Smaller values of $\varepsilon$ result in more splats being resampled. We observe that with excessively large values of $\varepsilon$, visible artifacts remain, whereas decreasing $\varepsilon$ eliminates these artifacts. For the teddy bear, $\varepsilon = 1.0$ leads to 6 resampling loops and takes 71.1 s, whereas $\varepsilon = 0.01$ leads to 9 resampling loops and takes 101.3 s. For the banana example, $\varepsilon = 0.1$ leads to 8 resampling loops and takes 82.2 s, whereas $\varepsilon = 0.01$ leads to 10 resampling loops and takes 244.9 s. In both cases, we find that taking $\varepsilon = 0.1$ is already satisfactory, and in general we remain below the 1 min 30 s limit.}
    \label{fig:resampling_epsilon}
\end{figure*}

\paragraph{Choice of $\varepsilon$ for resampling.}
The effect of the threshold hyperparameter $\varepsilon$ used in the resampling criterion is analyzed in \Fig~\ref{fig:resampling_epsilon}. The smaller $\varepsilon$ is, the more splats are resampled. This creates a trade-off: $\varepsilon$ should be small enough to remove artifacts through resampling, but large enough to prevent the number of splats from increasing excessively. \elieUpdate{The value of $\varepsilon$ also impacts the resampling time (1.5x to 3x slowdown when switching from $\varepsilon=0.1$ to $\varepsilon=0.01$). When resampling duration is not an issue (e.g., because one does it once for all before rendering an animated sequence), we find that} a grid search over the values $\{0.001, 0.01, 0.1, 1.0\}$ is sufficient for our examples. We start from $\varepsilon=1.0$ and decrease it until the artifacts disappear.


\end{document}

%% file: commands.tex
\usepackage[normalem]{ulem}

\definecolor{ElieColor}{RGB}{170,46,200}

\definecolor{NinaColor}{RGB}{255,165,0}

\newcommand{\elie}[1]{\textcolor{ElieColor}{É: #1}} %
\newcommand{\nina}[1]{\textcolor{NinaColor}{N: #1}} %
\newcommand{\elieDone}[1]{\elie{\sout{#1}}} %
\newcommand{\elieUpdate}[1]{{\color{ElieColor} #1}} %
\newcommand{\ninaUpdate}[1]{{\color{NinaColor} #1}}

\definecolor{SoutColor}{RGB}{200,145,0}

\renewcommand{\elie}[1]{}
\renewcommand{\elieDone}[1]{}
\renewcommand{\elieUpdate}[1]{#1}
\renewcommand{\nina}[1]{}
\renewcommand{\ninaUpdate}[1]{#1}

\newcommand\Sec{Sec.}
\newcommand\Fig{Fig.}
\newcommand\Eq{Eq.}
\newcommand\Tab{Tab.}
\newcommand\grad{\vec\nabla}
\newcommand\taureg{\tau_\text{reg}}
\DeclareMathOperator*{\argmin}{argmin}
\DeclareMathOperator*{\proj}{Proj} %
\DeclareMathOperator*{\sdist}{SignedDist}
\DeclareMathOperator*{\smoothstep}{smoothstep} %
\DeclareMathOperator*{\op}{Op} %
\DeclareMathOperator*{\softmax}{softmax}